%% file: 00-main.tex
\documentclass[sigconf]{acmart}

\AtBeginDocument{%
  \providecommand\BibTeX{{%
    \normalfont B\kern-0.5em{\scshape i\kern-0.25em b}\kern-0.8em\TeX}}}

\acmYear{2024}
\setcopyright{rightsretained}
\acmConference[CHI '24]{Proceedings of the CHI Conference on Human Factors in Computing Systems}{May 11--16, 2024}{Honolulu, HI, USA}
\acmBooktitle{Proceedings of the CHI Conference on Human Factors in Computing Systems (CHI '24), May 11--16, 2024, Honolulu, HI, USA}
\acmDOI{10.1145/3613904.3642289}
\acmISBN{979-8-4007-0330-0/24/05}

\usepackage{multirow}

\usepackage{enumitem}

\usepackage{subcaption}
\usepackage{xspace}

\newcommand{\low}{Low\xspace}
\newcommand{\med}{Medium\xspace}
\newcommand{\high}{High\xspace}

\usepackage{verbatim}

\newcommand{\new}[1]{#1}

\newcommand{\kr}[1]{}
\newcommand{\ta}[1]{}
\newcommand{\kl}[1]{}
\newcommand{\nascomment}[1]{}

\begin{document}

\title[Know Your Audience]{Know Your Audience: The benefits and pitfalls of generating plain language summaries beyond the ``general'' audience}

\author{Tal August}
\orcid{0000-0001-6726-4009}
\affiliation{%
  \institution{Allen Institute for AI}
  \city{Seattle}
  \state{Washington}
  \country{USA}
}

\author{Kyle Lo}
\orcid{0000-0002-1804-2853}
\affiliation{%
  \institution{Allen Institute for AI}
  \city{Seattle}
  \state{Washington}
  \country{USA}
}

\author{Noah A. Smith}
\orcid{0000-0002-2310-6380}
\affiliation{%
  \institution{University of Washington \& Allen Institute for AI}
  \city{Seattle}
  \state{Washington}
  \country{USA}
}

\author{Katharina Reinecke}
\orcid{0000-0001-7897-9325}
\affiliation{%
  \institution{University of Washington}
  \city{Seattle}
  \state{Washington}
  \country{USA}
}

\renewcommand{\shortauthors}{August et al.}

\begin{abstract}

Language models (LMs) show promise as tools for communicating science to the general public by simplifying and summarizing complex language. Because models can be prompted to generate text for a specific audience (e.g., college-educated adults), LMs might be used to create multiple versions of plain language summaries for people with different familiarities of scientific topics. However, it is not clear what the benefits and pitfalls of adaptive plain language are. When is simplifying necessary, what are the costs in doing so, and do these costs differ for readers with different background knowledge? Through three within-subjects studies in which we surface summaries for different envisioned audiences to participants of different backgrounds, we found that while simpler text led to the best reading experience for readers with little to no familiarity in a topic, high familiarity readers tended to ignore certain details in overly plain summaries (e.g., study limitations). Our work provides methods and guidance on ways of adapting plain language summaries beyond the single ``general'' audience.

\end{abstract}



\begin{teaserfigure}
\centering
    \includegraphics[width=0.75\textwidth]{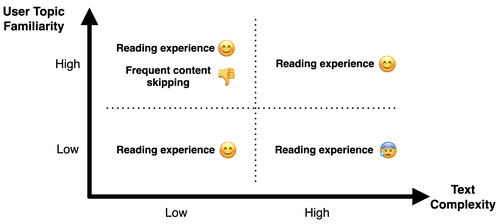}
    \caption{Simpler summaries were often the best reading experience for participants with little to no background in a scientific topic. However, readers with high topic familiarity, even those considered part of the general public (i.e., not a researcher), ignored more information in low complexity summaries while still reporting these simple summaries as equally engaging as high complexity ones. Our results provide guidance on generating plain language summaries for a wider range of general audiences.}
    \label{fig:teaser}
  \end{teaserfigure}
  
\maketitle

\input{01-introduction}

\input{02-RW}

\input{03-study1}

\input{04-study2}

\input{05-study3}

\input{08-discussion}

\begin{acks}
We thank Sarah Kahle for authoring the summaries and providing guidance on science communication practices and the Mechanical Turk annotators for their work on the project. We also thank the anonymous reviewers and members of the UWNLP and DUB community for their helpful feedback. This work was supported in part by the Office of Naval Research under MURI grant N00014-18-1-2670 and by a Twitch Research Fellowship. 
\end{acks}

\bibliographystyle{ACM-Reference-Format}
\bibliography{00-main}

\newpage
\appendix

\input{09-appendix}

\end{document}

%% file: 01-introduction.tex
\section{Introduction}

A rich body of work in HCI has shown that for many interfaces, one size does not fit all. Adapting interfaces to different users has the potential to improve usability \citep{Baughan2020KeepIS, Reinecke2011MOCCAA}, aesthetic judgements \citep{Lima2022AssessingTV, Mttus2015AestheticsOI}, and trust \citep{Lindgaard2011AnEO, Moshagen2010FacetsOV}. Increasingly, language styles, such as community language norms \citep{danescu2013no}, formality \citep{august2019pay}, and text complexity \citep{MartnezSilvagnoli2022OptimizingRA, August2022PaperPM} have been the focus of adaptable user interfaces. Work has shown that language styles can impact behavior in online experiments \citep{august2019pay}, counseling conversations \citep{althoff-etal-2016-large}, online communities \citep{danescu2013no}, and security interfaces \citep{stokes2023language}. This work has highlighted the benefits of adapting language to people with different backgrounds \citep{August2022PaperPM}.

With the rise of language models (LMs), interfaces promising adaptable language have progressed rapidly. Models like GPT-4 can ostensibly rewrite language for any reader by prompting the model to generate text for an envisioned audience or persona (e.g., a 5th grader) \citep{wu2023large, 10.1145/3526113.3545616, kirk2023personalisation}. This is especially enticing in scholarly and scientific communication, where language styles (e.g., medical jargon) can present major communication barriers \citep{Razack2021ArtificialIT}. Research has explored using models to adapt scientific papers for non-experts (referred to as general audience readers in this paper) \citep{August2022PaperPM, Guo2021AutomatedLL}, and paid services like Elicit,\footnote{\url{https://elicit.org/}} or Explainpaper\footnote{\url{https://www.explainpaper.com/}} promise to make scientific language easier to read and understand.

While adaptable language interfaces for communicating science are promising, it is not clear \new{when and how to adapt}. Most research showing that general audience readers respond positively to simplified language has focused on a single version of a simplified summary \new{and a single general audience} \citep{Guo2021AutomatedLL, Devaraj2021ParagraphlevelSO, Guo2023APPLSAM}. People have different knowledge and topic familiarity (\new{e.g., someone who has read popular science books on a subject compared to someone who has not}) that can impact how they respond to scientific information~\citep{nisbet2009s, Forzani2016IndividualDI, Bliss2019}, suggesting that a simplified summary may be good for some, while a more complex version may be advantageous for others. \new{However, no work has empirically shown this to be the case.} Further, simplified summaries usually convey less information \citep{august2020writing} and can unintentionally lead to people being overconfident in their understanding \citep{Scharrer2012TheSO}. In contexts where details are important, it may be important to preserve all information, even at the cost of longer or more complex text (e.g., a medical research paper \citep{August2022PaperPM}). \new{This gap in research is particularly important for developers of new interactive text interfaces \citep{Lo2023TheSR, August2022PaperPM} because it is currently not clear what the benefits and pitfalls of adaptive text are: when is simplifying necessary, what are the costs in doing so, and do these costs differ for readers with different background knowledge?}

Here we investigate how changes in scientific text affect the reading experience of general audience readers, \new{for the first time taking into account varying levels of complexity in the text and background topic familiarity of the reader}. We focus on scientific text complexity, defined as a combination of simple language and information content (\S \ref{sec:complexity}). We introduce three RQs to understand how changes in complexity and information content affect readers:

\begin{quote}
    \textbf{RQ1:} How do participants of different backgrounds respond to \textbf{human-written} scientific text at different complexity levels?
\end{quote}

\begin{quote}
    \textbf{RQ2:} How do participants of different backgrounds respond to \textbf{machine-generated} scientific text at different complexity levels?
\end{quote}

\begin{quote}
	\textbf{RQ3:} How do participants respond to generated scientific summaries at different complexities if they \textbf{report similar information}?
\end{quote}

We started with studying expert-written summaries (RQ1) to establish what benefit we might expect from using alternative complexity versions, assuming no interference from imperfect text generation tools. We followed up with two studies using machine-generated summaries. In study 2 we used generated summaries with no restriction on information content (RQ2), following prior work on generating scientific summaries for general audience readers \citep{guo2022cells, August2022GeneratingSD}. In study 3 we evaluated generated summaries that aimed to preserve information content in lower complexity summaries (i.e., explaining details rather than removing them) (RQ3). We ran within-subjects experiments on Mechanical Turk for each RQ (Study 1: $N=199$, Study 2: $N=191$, Study 3: $N = 203$) evaluating whether topic familiarity affected participants' response to summaries written or generated for different envisioned audiences at three levels of complexity.

We found that topic familiarity mattered for determining the ideal summary for a reader. While the lowest complexity summaries were generally better for people with minimal topical knowledge (illustrated in the lower left quadrant of Figure \ref{fig:teaser}), participants with more topic familiarity reported similar reading experiences across the three summary versions. Further, the lowest complexity summaries came with two costs to high familiarity participants. The first was that low complexity summaries in studies 1 and 2 removed details and reported on less information than high complexity summaries, shown with automatic and manual evaluations. This loss of information came with the benefit of improving the reading experience for low familiarity participants, but there was no benefit for high familiarity participants. The second, related cost was that high familiarity participants were more likely to skip sections of lower complexity summaries in all three studies (upper left quadrant of Figure \ref{fig:teaser}). The most commonly skipped text focused on a paper’s limitations, highlighting the risk that low complexity summaries have for high familiarity readers. 

\new{Our findings provide guidance on when and how to adapt scientific language to general audiences readers. Given our findings, we propose to only use the plainest language when an audience knows very little about a topic. In cases where audiences might have extensive background knowledge (even if they are not researchers themselves), language can be more complex---even drawn from the research paper---in order to convey more information and keep more knowledgeable audiences engaged (\S\ref{sec:studyOne} \& \ref{sec:studyTwo}). When it is vital to convey complete information, such as in a patient-clinician context, plain language that explains all information can still be beneficial even if it is much longer, but only to those with little knowledge of a scientific topic (\S\ref{sec:studyThree}). Our findings make the following contributions:}

\begin{enumerate}
    \item \new{\textbf{Shows the effect of text complexity on general audience readers of varying topic familiarity} (e.g., not comparing doctors and patients, but comparing different patients). We found that plain language summaries are better for those with little knowledge of a topic, and complex summaries, even those containing original scientific text, are better for those with more background knowledge.}

    \item \new{\textbf{Highlights the benefits and pitfalls of generating plain language summaries}. When plain language summaries matched a reader's background, readers had better reading experiences (e.g., were more engaged and had an easier time reading); however, plain language summaries often included less information and could lead to increased skipping when readers were more familiar in a topic. }

    \item \new{\textbf{Provides guidance on generating plain language for different audiences}. Science communicators and interface designers can use our findings and methodology (\S\ref{sec:studyTwoMaterials} \& \ref{sec:studyThreeMaterials}) to effectively provide multiple summaries of scientific findings to different people and build adaptive text interfaces. We discuss this guidance further in \S \ref{sec:guidanceDiscussion}.}
    
\end{enumerate}

\new{While LMs make it possible to generate language for a wide range of contexts and people, there are also risks of factually incorrect generations \citep{Maynez2020OnFA}. We discuss these risks in the context of science communication (\S\ref{sec:factuality}) and the need for expert oversight for generative systems (\S\ref{sec:discussion}). Our work illustrates ways for automated methods to assist human efforts in communicating scientific information to a wider range of people, going beyond a single general audience.}

%% file: 02-RW.tex
\section{Language Complexity}
\label{sec:complexity}
In this paper we define language complexity based on prior work in readability, plain language summarization, and science communication. Broadly we break down complexity along two dimensions: surface level, textual features of the language (referred to as ``plainness'' in this paper) and the information conveyed by the language (referred to as ``information content''). In this work we realize different language complexities by writing or generating summaries to different potential audiences (e.g., a high-school educated adult). 

In most science communication writing, both plainness and information content are varied to produce text suitable for different audiences. This joint variation is reflected in the guidelines for plain language summaries\footnote{\url{https://consumers.cochrane.org/PLEACS}} and in the strategies science writers use to communicate with interested publics \citep{august2020writing}. At the same time, these two dimensions have real-world constraints: there are situations in which technical words must be used to convey specific meaning, or where there is a desire to understand the majority of the details in the original scientific article, such as a patient reading a medical research paper or lab report \citep{Nunn2014LaySO, August2022PaperPM}. In studies 1 and 2, we allow plainness and information content to vary based on the intended audience (\S\ref{sec:studyOne} \& \S\ref{sec:studyTwo}). In study 3, we explicitly try to preserve information content by explaining rather than removing details from the high complexity summaries to evaluate the effect longer plain summaries have on readers of different backgrounds (\S\ref{sec:studyThree}).

\section{Related Work}

Below we cover additional prior work related to language personalization, plain language summaries for science communication, and augmented reading.

\subsection{Personalizing language}

There is a rich literature on adaptive interfaces and personalization in many domains, including website design \citep{Reinecke2014QuantifyingVP}, advertisement \citep{Urban2014,Hauser2009}, study recruitment \citep{august2018framing}, journalism \citep{Adar2017PersaLogPO}, and education \citep{Finkelstein2013TheEO, Dolog2003PersonalisationIE, Ogan2017ProficiencyAP}. Usually personalization focuses on adjusting visual elements, but work has also shown the benefit of adjusting language to different audiences. In the medical domain, \citet{Dimarco2007HealthDocCP} proposed HealthDoc, a system that generated personalized patient pamphlets according to patient demographic information, education, and health history. Prior work has found that such tailoring of patient pamphlets can improve health outcomes, including smoking behavior and future health complications~\citep{Strecher1994TheEO, Skinner1994PhysiciansRF, Marco2006AuthoringAG}. In journalism, \citet{Adar2017PersaLogPO} introduced PersaLog, a system for authoring personalized news articles. Articles authored using PersaLog presented alternative content (e.g., heat estimates for different areas) depending on user traits (e.g., a user’s location).  \citet{Finkelstein2013TheEO} showed that adjusting the dialect of a tutoring system could improve learning outcomes for children using African American English. Also in the education domain, work has shown that adjusting learning environments to learning styles or using personally-relevant examples can improve learning objectives \citep{KlanjaMilievi2011ELearningPB, DavisDorsey1991TheRO}.  Past work has also personalized generated news articles~\citep{Oh2020UnderstandingUP}, scientific definitions~\citep{Murthy2022ACCoRDAM}, recommended articles to read~\citep{Haruechaiyasak2008ArticleRB}, and the amount of text displayed in a website~\citep{Yu2010EnhancingWP}. 

Previous adaptive language-based interfaces have either relied on experts to author multiple versions of content~\citep{Adar2017PersaLogPO}, used rules and templates to automatically adjust content \citep{Oh2020UnderstandingUP, Dimarco2007HealthDocCP}, or focused on specialized populations (e.g., researchers \citep{Murthy2022ACCoRDAM}). Manually writing versions of text for each possible reader is infeasible, and rule-based approaches are brittle and only applicable to narrow content adaptation. In this paper we evaluate the feasibility of using modern NLP techniques to automatically generate multiple versions of text across a range of language complexities to communicate scientific information to different general audience readers.

\subsection{Plain language summaries}

Plain language summaries (PLS), also referred to as lay-summaries, patient summaries, or consumer summaries~\citep{Stoll2022PlainLS} are becoming an increasingly common method for communicating scientific findings with the public. \citet{10.7554/eLife.25411} surveyed ten organizations that produced plain language summaries, finding that while summaries might initially be intended for one audience (e.g., undergraduates), often other people would engage with the summaries~\citep{astrobites}.  

Studies have also explored how plain language summaries should be written based on empirical evidence from readers. \citet{Santesso2015AST} found that using structured headings and narrative flow improved comprehension compared to paragraphs of text explaining results. \citet{Ellen2014HealthSD} interviewed participants about their preferences for plain language summaries, finding that people prefer key message headings and bullets over paragraphs. \citet{MartnezSilvagnoli2022OptimizingRA} explored the preferences of summary text complexity, measured by automated readability formulas, across different age groups. They found that most people preferred a medium complexity, while the lowest complexity was viewed as too simple and the highest complexity as too hard. Other work has studied how to present numerical results in summaries \citep{Buljan2020FramingTN}, uncertainty in findings ~\citep{Alderdice2016DoCS} and how summaries compare to other methods of science outreach, such as infographics \citep{Buljan2018NoDI}, press releases \citep{Karacic2019LanguagesFD} and Wikipedia articles~\citep{Anzinger2020ComparativeUA}. In this paper we investigate if there is a benefit to adjusting the complexity of plain language summaries to different general audience readers. 

\subsection{Augmenting scientific reading}

New interaction techniques have augmented readers' process to improve understanding and engagement, especially for scientific text. \citet{Chaudhri2013InquireBA} introduced Inquire Biology, a biology textbook that allows students to view concept definitions and ask open-ended questions about information in the textbook. Work has also developed new interaction techniques for researchers reading papers, including surfacing definitions \citep{Head2021AugmentingSP}, searching over related work sections \citep{10.1145/3544548.3580841}, providing paper passages that answer natural language queries \citep{Zhao2020TalkTP} and navigating concepts within a paper \citep{Abekawa2016SideNoterSP, Jain2018ContentDE}. With the improved performance of LMs like GPT-3, 3.5, and 4 \citep{OpenAI2023GPT4TR}, there has been dramatic growth in augmented reading interfaces for scientific papers \citep{Lo2023TheSR}. For the general public, \citet{August2022PaperPM} introduced PaperPlain, a reading interface augmented with NLP to support general audience readers in approaching medical research papers. PaperPlain includes a curated set of key questions for guiding readers to the most important information in research papers. Augmented readers have also been released as products. Explainpaper\footnote{\url{https://www.explainpaper.com/}} is an LM-powered reading interface that allows users to ask questions over a paper and get simplified summaries.

Recent advances in NLP have also introduced automated methods to augment science communication~\citep{DemnerFushman2016AspiringTU, Wang2021PretrainedLM, Guo2021AutomatedLL}. ~\citet{Devaraj2021ParagraphlevelSO} introduced a dataset of plain language summaries for clinical topics and a trained model for simplifying medical information. \citet{laban-etal-2023-swipe} constructed a new dataset of simplification edits made on Wikipedia articles, \citet{ch2023medeasi} introduced a dataset of simplification edits for medical texts,  and \citet{Guo2023APPLSAM} introduced a new evaluation suite for plain language summarization. \citet{August2022GeneratingSD} introduced methods to generate definitions at different levels of complexity. \citet{Shaib2023SummarizingSA} evaluated simplified summaries of biomedical papers generated by GPT-3, finding that GPT-3 could simplify and summarize from single paper, but it struggled to synthesize information across multiple papers. 

Previous work for augmenting or generating scientific text either assumes there is a single ideal summary for all readers, or that adapting language to an individual reader is always useful. To our knowledge, no work has investigated if and when adaptation is important for scientific communication. \new{This is of particular importance to developers of augmented reading interfaces because it is currently not clear when augmentation or adaptation is necessary. For example, do all general audience readers need a reading interface to provide a plain language summary of a scientific paper? If so, should this summary look the same for everyone, or is there measurable improvement in reading experience if the summary matches the background of the reader? In this paper we investigate how general audience readers with different familiarity in a scientific topic respond to scientific text at different complexities to inform the development of augmented reading interfaces for scientific text. }

%% file: 03-study1.tex
\section{Study 1 -- Expert-written summaries}
\label{sec:studyOne}

Study 1 focused on expert-written summaries to establish what benefit we might expect from alternative complexity versions. The study answered our first research question: 

\begin{quote}
    \textbf{RQ1:} How do participants of different backgrounds respond to \textbf{human-written} scientific text at different complexity levels?
\end{quote}

Science writers adapt scientific language for general audiences. However, there is rarely a single general audience, and writers may use different strategies to engage different general audiences \citep{ranger2016kind, august2020writing}. Study 1 investigated how adjusting scientific language complexity affected people of different knowledge backgrounds.

\subsection{Method}

\begin{figure*}
    \centering
    \includegraphics[width=\textwidth]{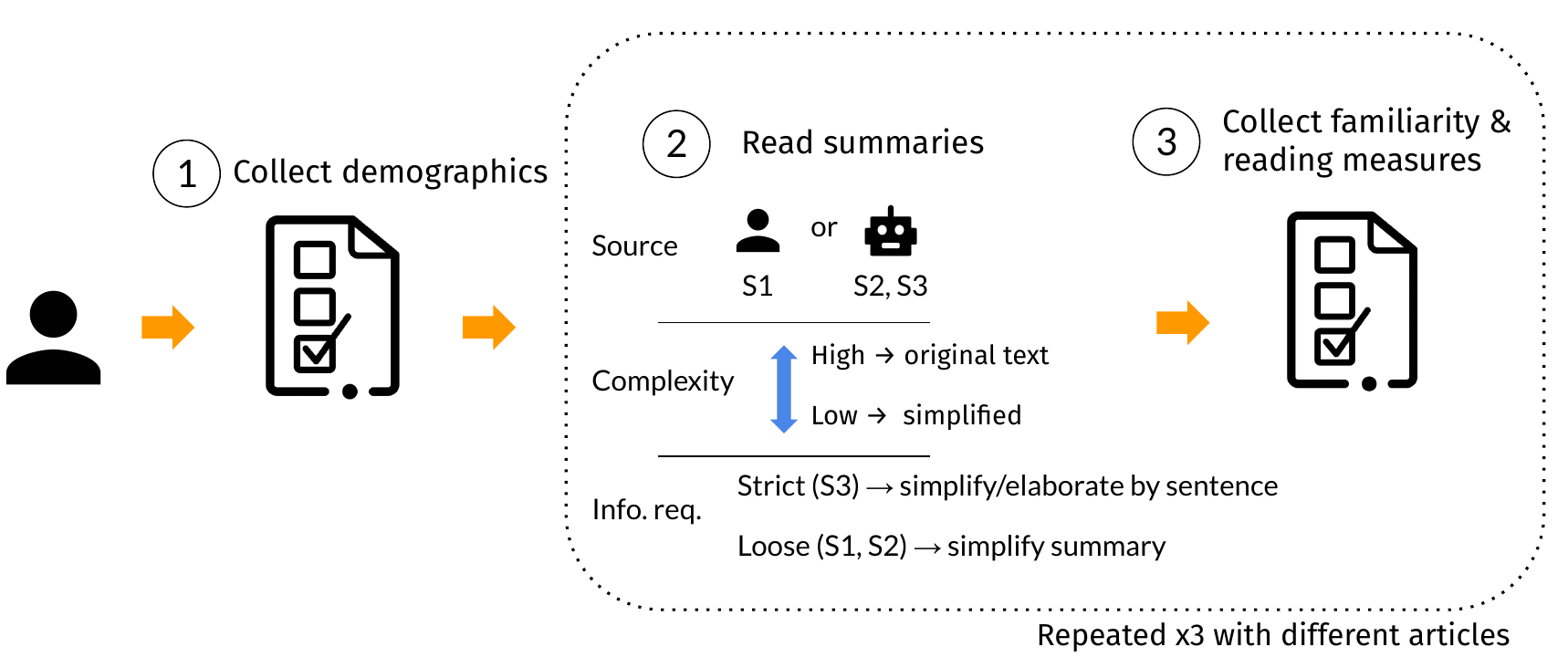}
    \caption{Flowchart of the study method, with shared features of all studies listed once. Ordering of summaries were randomized. \label{fig:studyMethod}}
\end{figure*}

\label{sec:methods}
The three studies shared the majority of their procedure, materials, participant recruitment, and analyses (Figure \ref{fig:studyMethod}). Below we report on the shared portions and those unique to study 1. Later, we report on differences in the methodology of studies 2 (\S\ref{sec:studyTwo}) and 3 (\S\ref{sec:studyThree}). 

\subsubsection{Procedure} 
\label{sec:procedure}

\begin{figure*}
    \centering
    \includegraphics[width=0.75\textwidth]{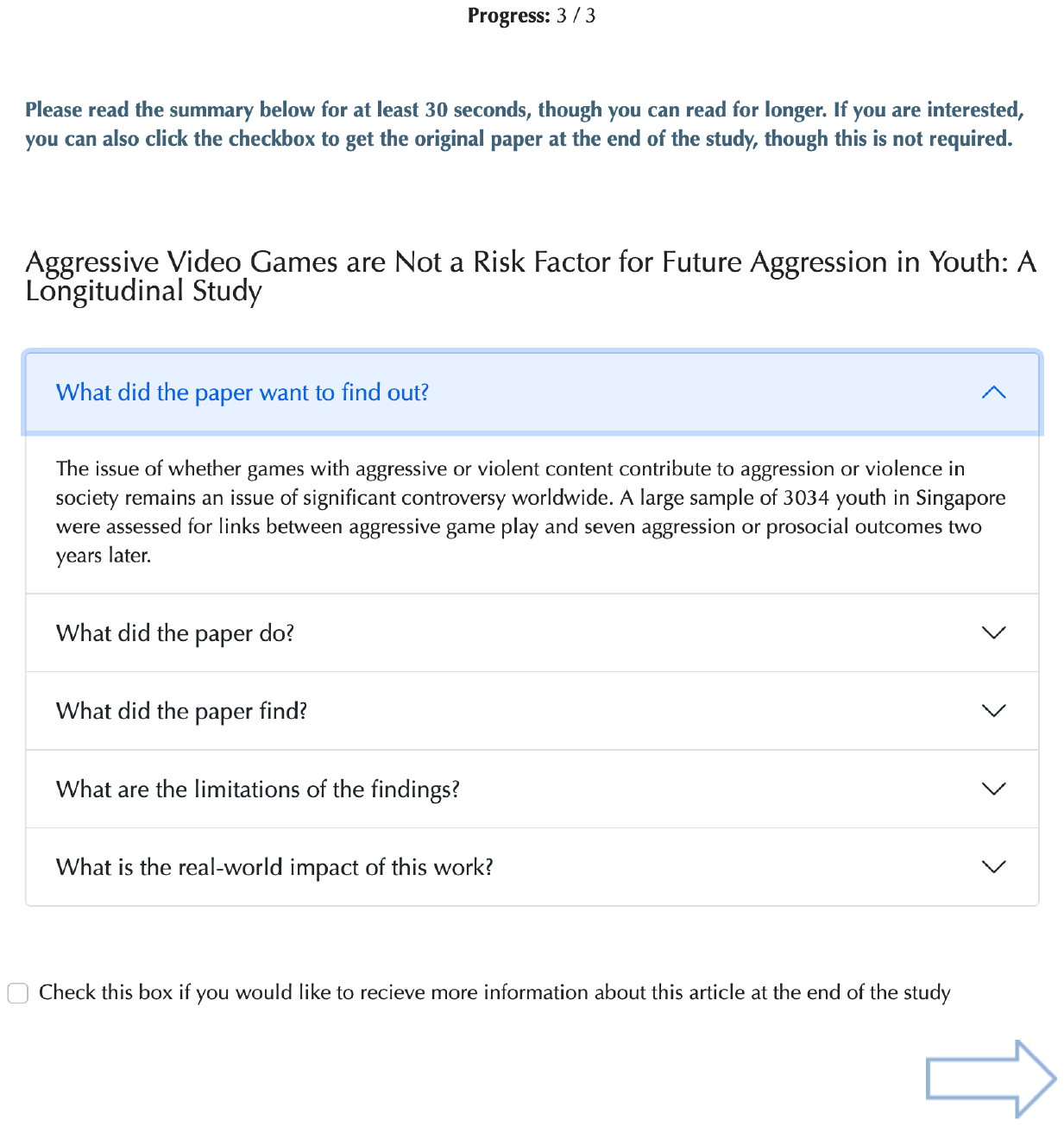}
    \caption{The study interface for reading the article summaries. The accordions started closed. \label{fig:studyScreenshot}}
\end{figure*}

Participants answered questions about their scientific background, read summaries of scientific papers at three levels of complexity, and answered questions about the summaries. At the start of each experiment, participants filled out a demographics questionnaire, including questions on their education, STEM experience, and interest in scientific subjects. After the demographic questionnaire, participants read three article summaries, described in \S\ref{sec:materials}. The articles and complexity levels were randomized. Each participant saw one of each complexity in random order. 

Summaries were broken down into sections answering key questions about the paper, following prior work showing that sections and headers were preferred by general audience readers~\citep{Santesso2015AST}. The key questions were based on prior work studying the key information that science communicators focus on in a paper~\cite{august2020writing, CochranePLS} and from questions general audience readers found useful to determine relevant information in research papers~\cite{August2022PaperPM}. Summaries were displayed as a title and a list of accordions (Figure \ref{fig:studyScreenshot}). Participants could open multiple accordions at once. The questions were:

\begin{enumerate}
    \item What did the paper want to find out?
    \item What did the paper do?
    \item What did the paper find?
    \item What are the limitations of the findings?
    \item What is the real world impact of this work?
\end{enumerate}

\new{Below the summary, participants could check a box requesting the original research paper. If participants checked this box, then a link to the paper was provided at the end of the study. } Participants were asked to read the summaries for at least 30 seconds, though they could read for as long as they wanted. If participants clicked the continue button before 30 seconds, they were prompted to read for at least 30 seconds. They could ignore this prompt by clicking the continue button again. Participants on average took 143 seconds per article (std=103 seconds) for study 1, 100 seconds (std=84) for study 2, and 137 seconds (std=78) for study 3. Participants then answered questions on their topic familiarity and reading experience. 

\subsubsection{Materials} 
\label{sec:materials}

\begin{table*}[]
\centering 
\small 
\begin{tabular}{p{25mm}p{20mm}p{90mm}}
\toprule
 \textbf{Source} & \textbf{Complexity} & \textbf{Summary}   \\ \midrule \midrule 
 
    & \high{} & These results demonstrate an unprecedented opportunity for development of these nanorgs as renewable sugar-free microbial factories for the production of biofuels and chemicals. 
    
    \\ 
    \cline{2-3}
    \\ 

    \multirow{2}{*}{Expert - Study 1} & \med{} &  This work is some of the first to examine the \textcolor{blue}{\textit{feasibility of interfacing nanoscale materials}} with living cells \dots{} which could have \textcolor{purple}{\textbf{broader implications for diagnostic and therapeutic applications of this technology.}} \\ \\

    & \low{} & This work is some of the first to be done investigating the \textcolor{blue}{\textit{possibility of using nanoscale materials}} inside living cells \ldots{} \textcolor{purple}{\textbf{which has far-ranging applications for medicine.}}  
    
    \\ 
    \cline{2-3}
    \\

    \multirow{2}{*}{Machine - Study 2} & \med{} &  The study found that nanorobots ... can be used to \textcolor{blue}{\textit{externally regulate the cellular function}} of living cells using \textcolor{purple}{\textbf{electromagnetic stimuli such as light, sound, or magnetic field.}} \\ \\

    & \low{} & This study found that nanorogs can be used ... to \textcolor{blue}{\textit{control living cells}} \textcolor{purple}{\textbf{using light, sound, or magnetic fields.}}

    \\ 
    \cline{2-3}
    \\ 

    \multirow{2}{*}{Machine - Study 3} & \med{} & This study shows that \textcolor{blue}{\textit{nanoscale organisms (nanorgs) can be developed into sustainable, sugar-free factories}}. \\ \\

     & \low{} & These findings show a new chance to create \textcolor{blue}{\textit{tiny organisms (called nanorgs)}} \ldots{} \textcolor{blue}{\textit{without using sugar, using sunlight in a way that can be reproduced on a larger scale}}. \\ \\

\bottomrule

\end{tabular}
\caption{\new{Examples of the summaries. These summaries were under the heading ``What are the real world impacts of the findings?'' for the same paper. } \textcolor{purple}{\textbf{Bolded purple text}} indicates examples of changes in information content between the summaries, and \textcolor{blue}{\textit{italicized blue text}} indicates changes in plainness. For study 2, there was no information restriction in generated summaries. In study 3, there was information restriction for generated summaries.} 
\label{tab:exampleSummaries}
\end{table*}
\paragraph{Article selection} 

We selected research papers that had public appeal by sampling papers posted and widely discussed in the large subreddit \textit{r/science} in 2019. We randomly sampled 10 papers posted on \textit{r/science} that contained a link to a research paper (as opposed to a press release or news article), and that had a score within the top 10\% of posts containing research papers. We used the \texttt{PSAW} Python PushShift API for accessing \textit{r/science}.\footnote{\url{https://psaw.readthedocs.io/en/latest/}} The papers ranged in topics from public policy to nanotechnology, reflecting the breadth of research papers posted and discussed on \textit{r/science}.

\paragraph{Authoring the summaries} 

An expert science writer with over 5 years of science communication experience crafted two versions of each summary. Each version was written for a different audience of a certain education level: a high school student or a college educated adult. In addition, the writer extracted sentences from the original paper to answer each key question. This constituted a third complexity aimed at other researchers. We defined these three complexity levels as \low{} (high school student), \med{} (college educated adult), and \high{} (researcher). Because the original paper text used a different voice than the other two versions, we lightly edited the \high{} version by changing ``we'' to ``the researchers.'' One author reviewed each summary version and provide feedback to the writer on language complexity between the three versions in four weekly meetings, as well as asynchronously with Google Docs. The rest of the authors reviewed the completed summaries to determine that each version was distinct from the others in language complexity. The writer was paid \$17.22 USD per hour. Table \ref{tab:exampleSummaries} provides examples of the summaries and Table \ref{tab:complexity_word_counts} lists word and sentence statistics for all summaries. All summaries are provided in the supplementary.  

\subsubsection{\new{Measuring language complexity}}
\label{sec:studyOneSimpleLangEval} 

\new{We additionally report on automated measures of complexity for each summary version in order to see how the generated summaries differ across complexity levels. Table~\ref{tab:complexity_word_counts} details the measures for each generated version. We report on three automated measures: uncommon words (i.e., English words outside the top 1,000 most common), function word count, and language model perplexity. While these measures do not capture all dimensions of complexity, they are measures for analyzing scientific complexity at scale used in prior work on adjusting language in science communication~\cite{August2022GeneratingSD, Guo2023APPLSAM}. Each measure is described in more detail in Appendix \ref{app:complexityMeasures}.}

\new{Table \ref{tab:complexity_word_counts} reports the results of the automated measures for all three studies. The \med{} and \low{} machine generated summaries in studies 2 and 3 had noticeable differences in average number of words, average proportion of uncommon English words outside the top 1,000, average proportion of function words, and language model perplexity. Compared to the expert written summaries, the generated summaries had more differences in the automated complexity measures, especially for generated text in study 2.}

\begin{table*}[t]
\centering \small
\begin{tabular}{llrrrrr}
\textbf{Source} & \textbf{Complexity} & \textbf{\# Words\textsubscript{std}} & \textbf{\# Sentences} & \textbf{Unc. Words $\uparrow$}  & \textbf{Func. words $\downarrow$}  & \textbf{Perplexity $\uparrow$}  \\ \hline

& \high{} & $483.10_{107.02}$ & $16.70_{4.03}$ & $0.55_{0.04}$ & $0.27_{0.03}$ & $94.60_{30.91}$ \\ \\
\multirow{2}{*}{Expert - Study 1} & \med{} & $369.60_{82.56}$ & $12.10_{1.97}$ & $0.47_{0.05}$ & $0.31_{0.02}$ & $60.08_{14.84}$ \\
& \low{} & $358.90_{98.92}$ & $11.50_{3.21}$ & $0.43_{0.05}$ & $0.32_{0.02}$ & $53.68_{9.96}$ \\ \\
\multirow{2}{*}{Machine - Study 2} & \med{} & $529.20_{182.48}$ & $20.80_{7.05}$ & $0.51_{0.04}$ & $0.31_{0.03}$ & $64.14_{24.16}$ \\
& \low{} & $259.00_{43.42}$ & $13.20_{1.75}$ & $0.28_{0.05}$ & $0.36_{0.03}$ & $23.92_{5.71}$ \\ \\
\multirow{2}{*}{Machine - Study 3} & \med{} & $878.90_{212.81}$ & $31.00_{8.06}$ & $0.48_{0.02}$ & $0.34_{0.02}$ & $46.26_{9.90}$ \\
& \low{} & $1005.00_{273.63}$ & $37.70_{9.91}$ & $0.37_{0.03}$ & $0.37_{0.02}$ & $34.40_{4.83}$ \\

\end{tabular}
\caption{\new{Average number of words and sentences, along with differences in automated complexity measures between in each summary version. For study 2, there was no information restriction in generated summaries. In study 3, the summaries were generally longer because they included more details from the \high{} summaries (i.e., they had stricter information requirements). Arrows denote expected increase ($\uparrow$) or decrease ($\downarrow$) in measure as complexity increases.}}
\label{tab:complexity_word_counts}
\end{table*}

\subsubsection{Participants}
\label{sec:participants}

We recruited participants on Amazon Mechanical Turk with the slogan, ``Read about interesting scientific findings and answer questions about your experience.'' Participants were paid \$2.50. Participants were required to have completed over 1,000 HITs with a minimum approval rating of 95\% and be US-based. For studies 1 and 3, participants were required to be master Turkers. This study was approved by our institution's IRB. \new{We removed participants whose native language was not English (1 in study 1, 2 in study 2, and 3 in study 3)} and who indicated in a final self-report survey that they had technical difficulties or were cheating  (1, 12, and 0, respectively). After removal, we had 199 participants for study 1, 191 for study 2, and 203 for study 3.  Table \ref{tab:demographics-and-familiarity} lists demographics and topic familiarity.

Extrinsic motivations like payment can lead participants to maximize pay at the expense of data quality (e.g., by rushing through a study \citep{august2019pay, 10.1145/3022198.3026339}). Studies 1 and 3 used Master Turkers, who have been shown to provide data quality equivalent to intrinsically motivated participants (e.g., participants motivated by supporting science) \citep{10.1145/3022198.3026339}. After finding comparable results between master and non-master workers in a study 2 pilot, we did not include the masters requirement for study 2. However, we did have to remove more participants who had reported cheating during study 2. 

While participants might have behaved differently (e.g., skipped less sections, \S\ref{sec:readingMeasures}) if they were interested in the summaries for their own sake, we did not expect this to bias differences across complexity versions due to the within-subjects nature of the studies. Considering that prior work studying general audience readers of scientific articles has found that readers may skip parts of an article \citep{Conlen2019CaptureA}, we are excited to investigate how our findings generalize to readers motivated simply by interest in a topic.

\begin{table*}[h]
    \begin{subtable}[h]{0.50\textwidth}
        \centering
        \begin{tabular}{lllll}
        \toprule
         &  & Study 1 & Study 2 & Study 3\\
         \midrule
        \multirow{8}{*}{Age} & 0-19 & 0  & 0 & 0\\
         & 20-29 & 14& 49 & 9\\
         & 30-39 & 68 & 87& 76\\
         & 40-49 & 71 & 32& 57\\
         & 50-59 & 29 & 18 & 29\\
         & 60-69 & 14 & 4 & 21\\
         & 70-79 & 3 & 1 & 2\\
         & 80+ & 0 & 0 & 0\\
         \midrule
        \multirow{3}{*}{Gender} & Male & 98 & 96& 93\\
        & Female & 99 & 95& 109 \\
        & Prefer not to answer & 2 & 0& 4 \\ 
         \midrule
        \multirow{4}{*}{Education} & Pre-high school & 0 & 1 & 0 \\
         & High school & 58 & 30 & 48\\
         & College  & 117 & 114& 137\\
         & Graduate school & 19 & 40& 20\\
         & Professional school & 5 & 6 & 1\\
         \midrule
        \multirow{4}{*}{\begin{tabular}[c]{@{}l@{}}\# STEM courses\\ after high school\end{tabular}} & 0 & 36  &  21 & 36\\
         & 1--3 & 89 & 93 & 104\\
         & 4--6 & 41& 57 & 32\\
         & 7--10 & 11 & 9& 10\\
         & $\ge$11 & 22& 11& 21\\
         \bottomrule
        \end{tabular}
       \caption{Participant demographics}
       \label{tab:demographics}
    \end{subtable}
    \hfill
    \begin{subtable}[h]{0.35\textwidth}
        \centering
        \begin{tabular}{clll}
        \toprule
         Familiarity & Study 1 & Study 2 & Study 3 \\
         \midrule
         1 & 359& 150& 297\\
         2  & 115& 72& 134\\
         3 & 97 &  132 & 134\\
         4 & 26 & 165 & 39\\
         5 & 0  & 54 & 5\\
         \midrule
         Total & 597 & 573 & 609 \\
        \bottomrule
        \end{tabular}
        \caption{Topic familiarity based on question ``How familiar are you with the topic of this article?'' 1=``I have never heard about this topic before'', and 5=``I have written research papers on this topic.''} 
        \label{tab:familiarity}
     \end{subtable}
     \caption{Participant demographics (a) and topic familiarity (b) for all studies}
    \label{tab:demographics-and-familiarity}
\end{table*}

\subsubsection{Measures}
\label{sec:readingMeasures}

\paragraph{Topic familiarity} 

After each summary, participants rated their familiarity with the article’s topic on a 1---5 Likert-style scale based on the question: ``How familiar are you with the topic of this article?''\footnote{Because participants were only ever presented summaries, not the original paper, in the study the summaries were referred to as `articles.'} with 1 being ``I have never heard about this topic before'' and 5 being ``I have written research papers on this topic.'' Table~\ref{tab:familiarity} details the topic familiarity ratings for the three studies. 

\paragraph{Reading experience ratings}

We collected subjective ratings to understand how the different complexity levels affected participants' reading experience. Participants completed the ratings after reading each summary. All ratings were based on a 1--5 Likert-style scale. These included:

\begin{enumerate}
    \item \textbf{Reading ease}: Participants rated their reading difficulty based on the question: ``How easy was it for you to read the article?''
    \item \textbf{Understanding}: Participants rated their confidence understanding the summary based on the question: ``How confident do you feel in your understanding of the article?''
    \item \textbf{Interest}: Participants rated how interesting they found a summary based on the question: ``How interesting did you find the article?''
    \item \textbf{Value}: Participants rated how valuable they found the information in the summary based on the question: ``How much would you agree that this article contained valuable information?''

\end{enumerate}

\paragraph{Skipped sections}

We analyzed how many summary sections participants skipped in each complexity condition. As described in \S\ref{sec:procedure}, each summary was made up of five accordian drop-downs that participants could open. Each accordian section began closed. Participants were not instructed to open all sections. To determine which sections were opened, we logged click events for each accordian section.

\paragraph{\new{Requested articles}}

\new{A primary goal of science communication is to encourage audiences to engage further with science \citep{nisbet2009s}. We capture the potential for increased engagement with science by analysing how likely participants were to request the original scientific article after reading a summary. }

\subsubsection{Analysis} 
\label{sec:analysis}

We compared measures across the complexity versions using linear mixed-effects models (LMMs). \new{LMMs are commonly used to analyze data in which the same participant provides multiple, possibly correlated, measurements, referred to as repeated measures~\citep{Lindstrom1990NonlinearME} and have been used as an analysis tool in the behavioral sciences~\cite{Cudeck1996MixedeffectsMI} and human-computer interaction~\cite{Hearst2020AnEO, Head2021AugmentingSP}. }

We fit a model for each reading experience rating, number of skipped sections, and original article requests. Each model contained fixed effects for the complexity version, topic familiarity, an interaction term for familiarity and complexity, and random effects for paper and participant IDs. We conducted post-hoc two-sided $t$-tests for pairwise comparisons to examine the differences in measures between pairs of complexity levels estimated by the linear mixed effects models. \new{These pairwise comparisons reveal not only what differences between measures are significant, but the estimated differences $d$ between measures. Because $d$ is estimated by the linear mixed-effects model, it represents the expected difference in some measure (e.g., reading ease), when controlling for the participant and paper random effects in the model. For example, if the estimated difference $d^{\textrm{Low} - \textrm{High}}$ in reading ease between two complexity options \low{} and \high{} is 0.894, we can interpret this difference as participants rated the \low{} complexity, on average, 0.894 points higher for reading ease (out of 5) compared to the \high{} complexity when controlling for participant and paper. We report these differences to provide further intuition about the effect of different complexity levels. We also include effect sizes, calculated using Cohen's $d$ and denoted $SMD$ for standardized mean difference, as an additional measure of effect beyond the estimated pairwise difference. }

\new{The reading experience measures used Likert-style scales, making parametric tests potentially not appropriate, we report analogous non-parametric tests in Appendix \ref{app:ordinalRegression}, which yield similar $p$-values and findings.} For these analyses we use the \textsc{pymer4} Python package for fitting the models and pairwise comparisons. All $t$-tests were corrected from multiple hypotheses using the Holm-Bonferroni correction. The analysis was equivalent for the three studies. \new{We report all pairwise differences and test statistics in Appendix \ref{app:testStats}.}

\subsection{Results}
\label{sec:resultsOne}

\new{Table \ref{tab:pairwise-contrasts-1} in the appendix lists all pairwise differences.}

\subsubsection{Reading experience measures}
\label{sec:resultsExperienceStudyOne}

\begin{figure*}
     \centering
     \begin{subfigure}[]{\textwidth}
         \centering
         \includegraphics[width=\textwidth]{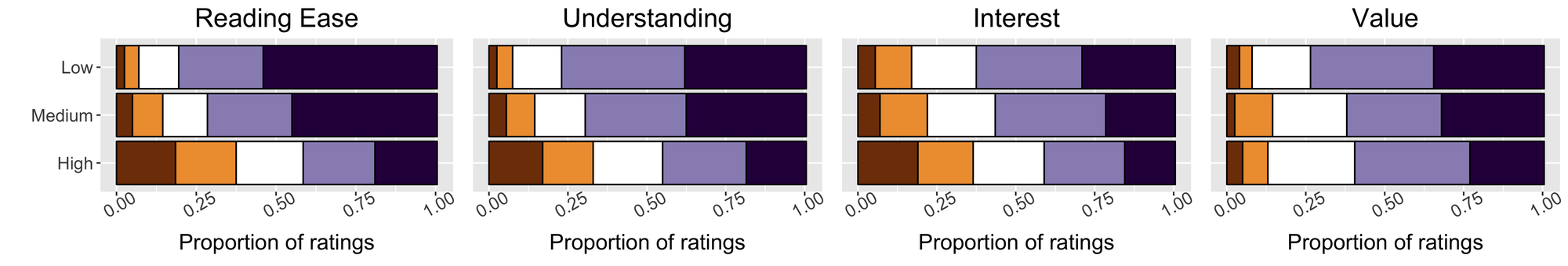}
            \caption{Study 1 with expert-written summaries}
    \label{fig:subjective_ratings_study1}
     \end{subfigure}
     \hfill
     \begin{subfigure}[]{\textwidth}
         \centering
         \includegraphics[width=\textwidth]{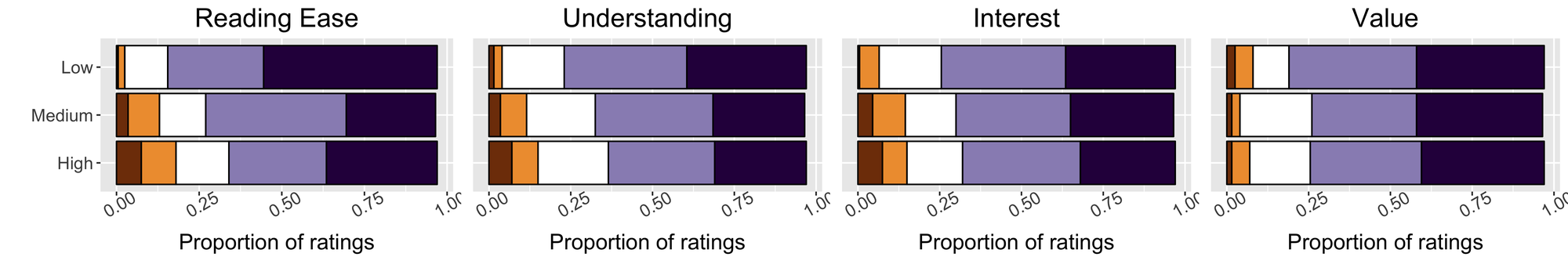}
    \caption{Study 2 with machine-generated summaries and no information restriction.}
    \label{fig:subjective_ratings_study2}
     \end{subfigure}
     \hfill
     \begin{subfigure}[]{\textwidth}
         \centering
         \includegraphics[width=\textwidth]{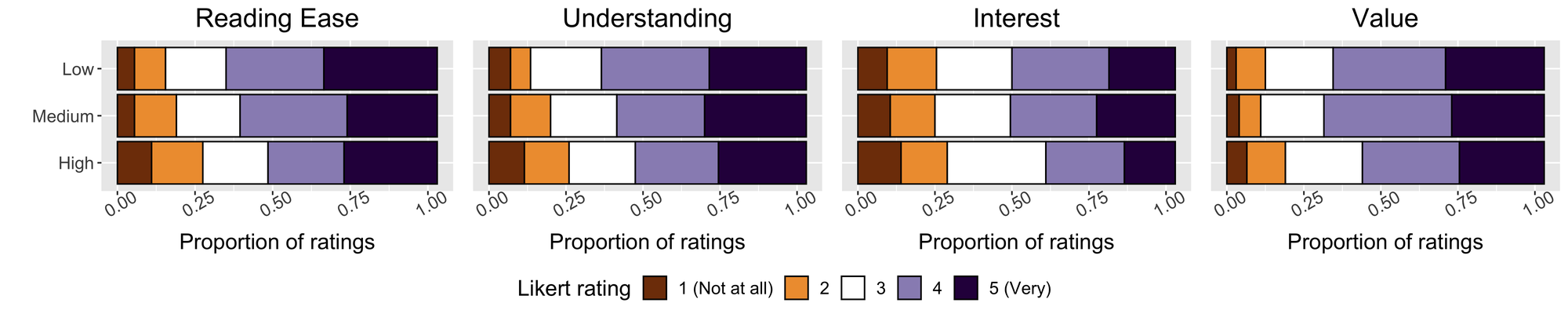}
    \caption{Study 3 with machine-generated summaries and information restriction.}
    \label{fig:subjective_ratings_study3}
     \end{subfigure}
        \caption{Distribution of ratings  for each subjective reading experience measure across complexity levels. The ratings were based on the following questions: Reading ease: ``How easy was it for you to read the article?'', Understanding: ``How confident do you feel in your understanding of the article?'', Interest: ``How interesting did you find the article?'', Value: ``How much would you agree that this article contained valuable information?'' Notice the greater number of high ratings (purple) and fewer low ratings (orange) as participants are presented with less complex summaries.}
        \label{fig:subjective_ratings_all}
\end{figure*}

\begin{figure*}
     \centering
     \includegraphics[width=\textwidth]{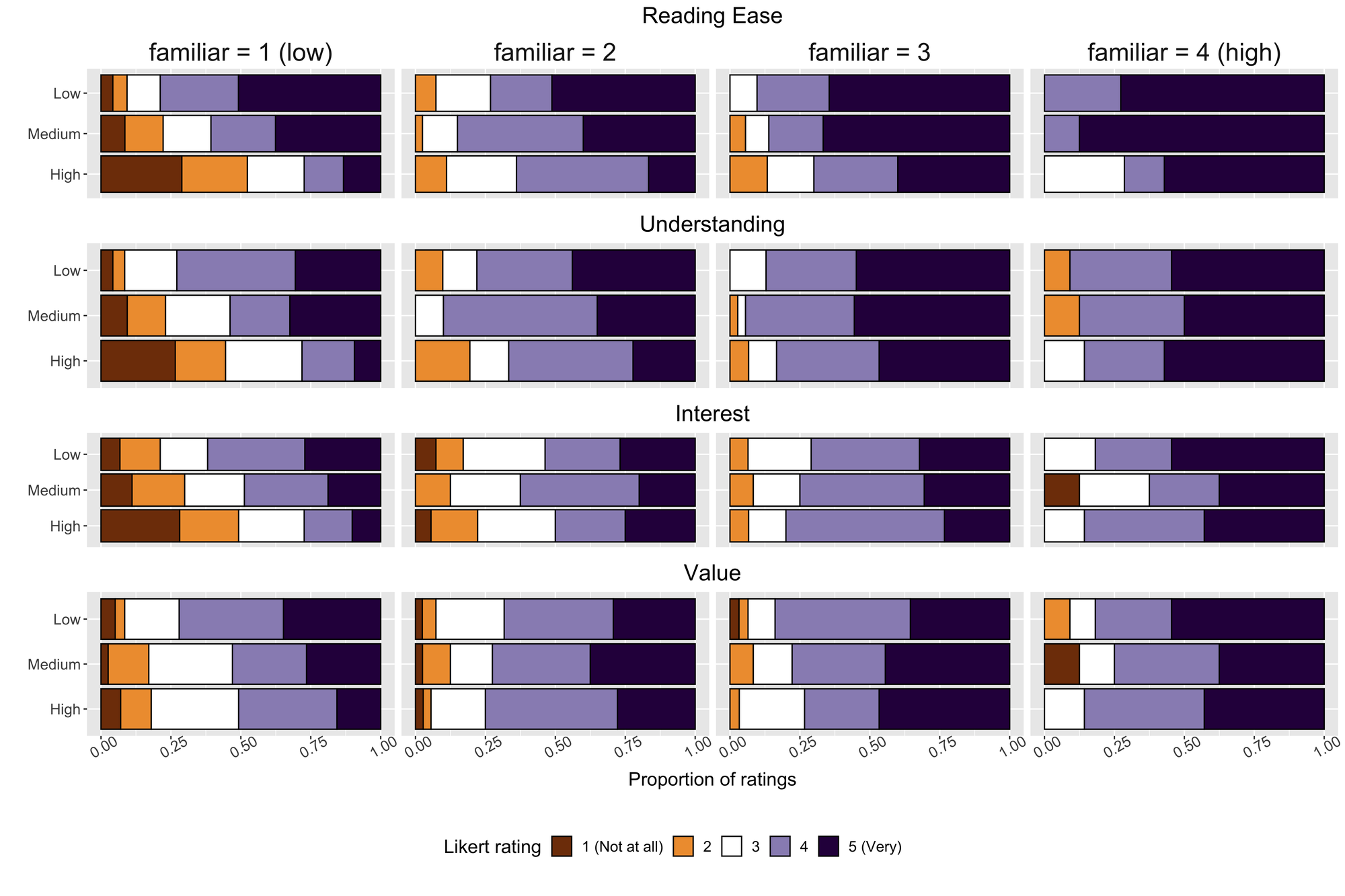}
        \caption{Distribution of ratings  for each reading experience measure across complexity and participant topic familiarity for study 1 (expert written summaries).}
    \label{fig:complexityInteractionsStudy1}
\end{figure*}

Figure~\ref{fig:subjective_ratings_study1} plots all participants' ratings across summary complexities for study 1.  Overall participants found the \low{} summaries most appealing. Across all measures there is a greater number of high ratings and fewer low ratings as participants are presented with less complex summaries. Compared to the \high{} summaries, participants rated \low{} summaries as significantly easier to read ($d_{\textit{ease}}=0.893$, $p<0.0001$, $SMD=0.99$), understand ($d_{\textit{understand}}=0.589$, $p<0.0001$, $SMD=0.77$), and more interesting ($d_{\textit{interest}}=0.381$, $p=0.018$, $SMD=0.55$). Participants also rated the \med{} summaries as significantly easier to read and were more confident in their understanding compared to the \high{} summaries ($d_{\textit{ease}}=0.653$, $p<0.0001$, $SMD=0.71$; $d_{\textit{understand}}=0.400$, $p=0.006$, $SMD=0.59$).

Topic familiarity was a strong indicator of reading experience measures and interacted with summary complexity. Looking at Figure ~\ref{fig:complexityInteractionsStudy1}, as familiarity increased, ratings across all metrics and complexity levels generally went up (i.e., the orange bars shrink while the dark purple bars grow). Also apparent in Figure~\ref{fig:complexityInteractionsStudy1}: at low familiarity, rating distribution are most different across the complexity levels. As familiarity increases, though, there were fewer low ratings and more high ratings for all complexity levels. This effect was also illustrated in the linear mixed effect models. Participants who rated their familiarity with a summary's topic lowest (1 on a scale of 1---5) rated the \low{} summaries as being significantly easier to read, understand, more interesting, and containing more valuable information compared to the \high{} summaries in study 1 ($d_{\textit{ease}}=1.490$, $SMD=1.27$; $d_{\textit{understand}}=1.160$, $SMD=1.07$; $d_{\textit{interest}}=0.943$, $SMD=0.80$ ; $d_{\textit{value}}=0.509$, $SMD=0.49$; $p<0.0001$ for all comparisons). Participants who were most familiar with the summary's topic, though, rated \high{} complexity summaries as similarly easy to read and understand, and equally interesting and valuable as \low{} and \med{} summaries. \new{Table \ref{tab:pairwise-contrasts-1} in the appendix lists all pairwise differences.}

\subsubsection{Skipped sections}
\label{sec:resultsUnclickedOne}

Participants on average skipped $0.113$ (std = $0.536$) sections (out of 5). Skipped sections were lowest for the \high{} summaries (mean=$0.060$, std=$0.327$) compared to the \low{} (mean = $0.129$, std = $0.559$) and \med{} (mean = $0.149$ std = $0.661$) summaries. Topic familiarity mattered for determining number of skipped sections. Participants who rated their topic familiarity highest (4 on a 1---5 scale), clicked on significantly fewer sections in the \low{} summaries compared to the \high{} summaries ($d_{unclicked}=0.682$, $p=0.008$, $SMD=0.68$). \new{Table \ref{tab:pairwise-contrasts-1} in the appendix lists all pairwise differences between skipped sections.} Across all studies, the most common section skipped by participants was the paper's limitations (``What are the limitations of the findings?'', 25\% of skipped sections), the least common section was the paper's goals (``What did the paper want to find out?'', 13\%). 

\subsubsection{\new{Original article requests}}
\label{sec:resultsOriginalArticle}

\new{Participants on average requested the original article 14.7\% of the time. Requests were roughly similar across the complexity levels (\low{}: mean=14.5\%, \med{}: mean = 15.5\%, \high{}: mean=14.0\%). Topic familiarity affected how likely participants were to request the original article depending on complexity level.  Participants with the second lowest familiarity (2 out of 5) requested the original article significantly less often in the \low{} summaries compared to the \high{} summaries ($d_{requests}=-0.184$, $p=0.007$, $SMD=-0.47$). Table \ref{tab:pairwise-contrasts-1} in the appendix lists all differences. }

%% file: 04-study2.tex
\section{Study 2 - Machine-Generated Summaries with no restriction on information content}
\label{sec:studyTwo}

The results from study 1 suggest that low complexity summaries are best for low familiarity participants, while high familiarity participants were more likely to skip sections in low complexity summaries. We were curios if we would see similar differences in complexity preference with machine-generated summaries. We therefore conducted study 2, answering our second research question: 

\begin{quote}
    \textbf{RQ2:} How do participants of different backgrounds respond to \textbf{machine-generated} scientific text at different complexity levels?
\end{quote}

There are methods to automatically adjust generated language complexity \citep{August2022GeneratingSD}, but no work has explored the interaction of generated language complexity and participant background knowledge. Here we follow prior work on automated plain language summarization and allow generated text to vary information content freely \citep{guo2022cells, laban-etal-2023-swipe}. In study 3 we explore methods to preserve information through all complexity levels (\S\ref{sec:studyThree}).  

\subsection{Method}
\label{sec:studyTwoMethod}
Below we describe generating summaries for study 2 and assessing their factuality. Please refer to \S\ref{sec:methods} for shared methodology of studies 1, 2, and 3.

\subsubsection{Materials}
\label{sec:studyTwoMaterials}

\paragraph{Generating the summaries}

We generated summaries at different complexities in a two step process. In the first step, we generated candidate summaries using GPT-3. GPT-3 is a language model commonly used in generation tasks, including plain language summarization \citep{Brown2020LanguageMA}. We adapted a preset prompt for GPT-3 to generate summaries with varying complexity. The original prompt was ``Summarize this for a second-grade student: [TEXT]'' Our adapted prompts for GPT-3 were $14$ alternate prompts, from ``first-grade student'' to ``twelfth-grade student'', along  with ``college student'' and ``college-educated adult.'' We used GPT-3 (\textsc{davinci-003}) in July 2022. with temperature set to 0.3 and the rest of the parameters set to default OpenAI API settings. At the time we ran this study, more sophisticated systems like ChatGPT had not been released. We investigate more sophisticated models (i.e., GPT-3.5 Turbo) in Study 3 (\S\ref{sec:studyThreeMaterials}).  

Because GPT-3 was not designed to explicitly vary text complexity, we additionally used the complexity ranker from \citet{August2022GeneratingSD} to rank the GPT-3 generations on a gradient of complexity. The complexity ranker was a linear discriminator trained to classify scientific text as either from a news article or research paper. The ranker used features shown to be predictive of reading difficulty in scientific language, including technical word occurrences, proportion of function words, and text length \citep{August2022GeneratingSD}. After scoring each generation for complexity, we selected the generation with the highest and lowest score for the \low{} and \med{} versions.  For the \high{} summaries, we used the original sentences extracted from the paper by the writer in \S\ref{sec:materials}. More details on the GPT-3 generations are in Appendix \ref{app:generatingSummaries}.

\subsubsection{Assessing factuality in generated summaries}
\label{sec:factuality}

A major limitation of language models is that they can generate text with meaning that was not part of the original input \citep{Maynez2020OnFA}, referred to as hallucinations \citep{Maynez2020OnFA, Goyal2021AnnotatingAM}. While there are methods for reducing hallucinations or encouraging factuality \citep{Gabriel2021DiscourseUA, Lu2021NeuroLogicD, laban-etal-2022-summac}, no automated method guarantees factual accuracy or fidelity to original text. In the context of science communication, such hallucinations can risk confusing or, worse, misinforming readers. A reader might trust a hallucinated result opposite to what was reported in the original paper~\citep{Devaraj2022EvaluatingFI}, or be so confused by the contradictory evidence as to lose trust in the research. 

Because of these risks, we advocate for NLP systems to be used in conjunction with experts. Plain language summaries are often written by researchers, editors, or science writers~\citep{Stoll2022PlainLS, 10.7554/eLife.25411}. Authors could generate multiple versions of a summary and then verify  factual accuracy. In this way, we could  lessen the workload of writing plain language summaries, make summaries adaptable to different audiences, and protect against factually incorrect generations. 

In the context of study 2 and 3, one author selected generations that did not contain factually incorrect information, acting as the expert for checking generated summaries before publishing. In study 2, out of 120 generated summaries (6 sections including the title $\times$ 10 papers $\times$ 2 complexities), 14 generations contained incorrect information. In all 14 cases, a replacement was found by selecting from at most 6 alternative generations. The average number of generations the author looked at to find a replacement was $2.36$. For study 3, while there were generations that were ill-formed (e.g., the model asking for clarification on an acronym) there were no factually incorrect generations.  This difference in factuality might be due to improvements between GPT-3 (used in study 2) and GPT-3.5 (used in study 3).\footnote{Because GPT-3.5 is a proprietary system, the full details of which have not been disclosed, we cannot be certain about whether or how factuality was improved.} Appendix~\ref{app:factuality} contains more information on hallucinations in our generated summaries.

\subsection{Results}
\label{sec:studyTwoResults}

\subsubsection{Reading experience measures}

\begin{figure*}
     \centering
     \includegraphics[width=\textwidth]{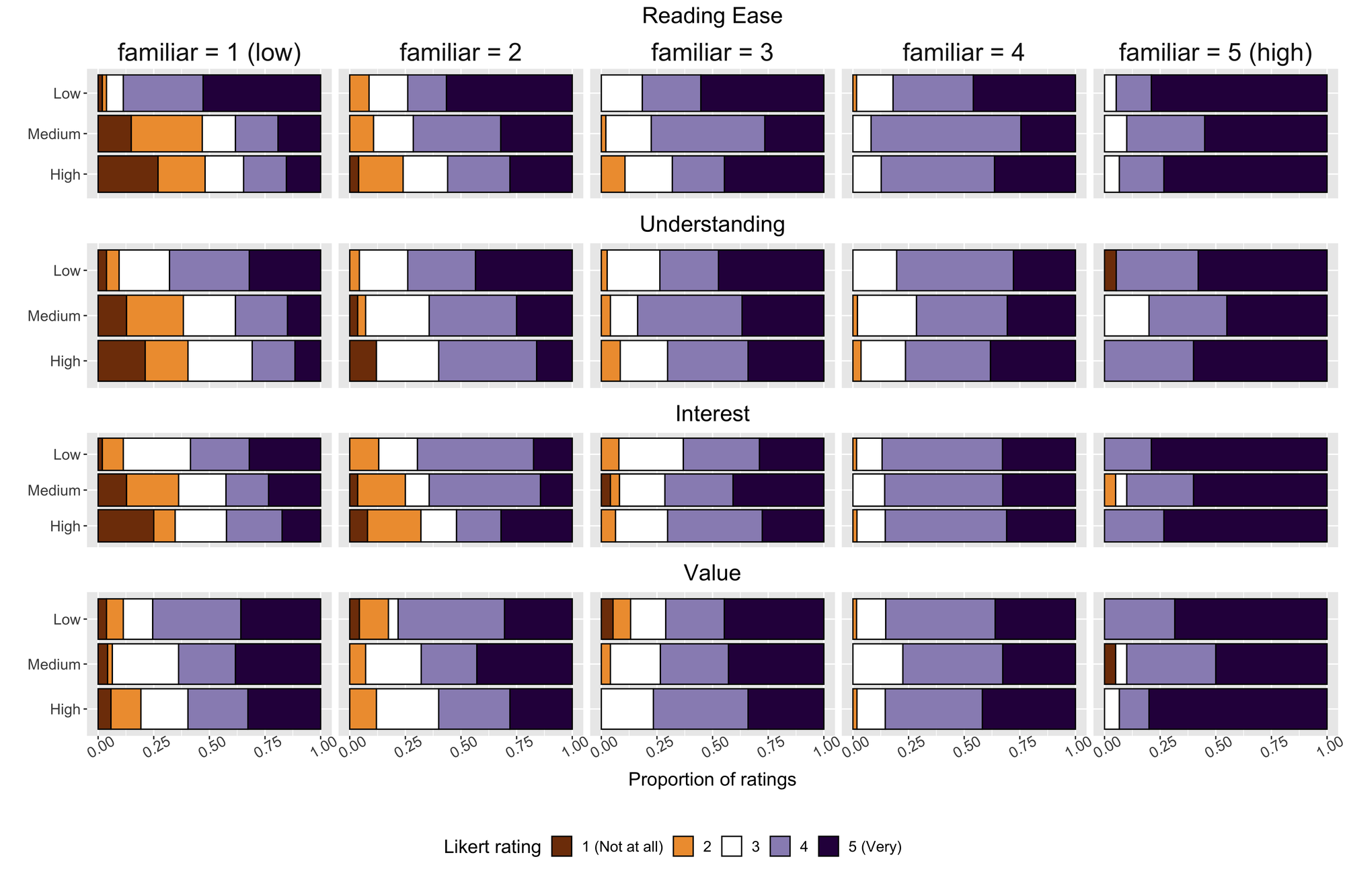}
        \caption{Distribution of ratings  for each reading experience measure across complexity and participant topic familiarity for study 2 (machine-generated summaries and no information restriction).}
    \label{fig:complexityInteractionsStudy2}
\end{figure*}

Similar to study 1, participants in study 2 rated \low{} summaries as significantly easier to read ($d_{\textit{ease}}=0.535$, $p<0.0001$, $SMD=0.56$) and understand ($d_{\textit{understand}}=0.323$, $p=0.001$, $SMD=0.38$) than the \high{} summaries (Figure~\ref{fig:subjective_ratings_study2}). However, we observed two different results in this second study. First, while study 1 participants found \med{} summaries significantly easier to read and understand than \high{} summaries, study 2 participants did not. Second, while study 1 participants did not rate the \low{} and \med{} summaries as significantly different, study 2 participants \emph{did} rate \low{} summaries as significantly easier to read and understand than \med{} summaries ($d_{\textit{ease}}=0.472$, $p<0.0001$, $SMD=0.53$; $d_{\textit{understand}}=0.279$, $p=0.004$, $SMD=0.29$).

Topic familiarity again interacted with complexity to equalize reading experience measures. Similar to study 1, participants with the lowest familiarity of a summary's topic rated the \low{} summaries as being significantly easier to read, understand, more interesting, and containing more valuable information compared to the \high{} summaries ($d_{\textit{ease}}=1.642$, $SMD=1.34$ $d_{\textit{understand}}=1.103$, $SMD=0.94$  $d_{\textit{interest}}=0.909$, $SMD=0.62$  $p<0.0001$; $d_{\textit{value}}=0.407$, $p=0.031$, $SMD=0.25$ ). In contrast, participants with the highest familiarity (5 on a 1--5 scale) rated their reading experience similarly between the complexity versions. Figure~\ref{fig:complexityInteractionsStudy2} plots ratings. 

\subsubsection{Skipped sections}

Participants on average skipped $0.785$ (std = $1.621$) sections in study 2. While the overall rate of skipped sections was higher than for study 1, the trend of more skipped sections for lower complexity summaries held. Skipped sections were lowest for the \high{} summaries (mean=$0.749$, std=$1.543$) compared to the \low{} (mean=$0.849$, std=$1.710$) and \med{} (mean=$0.759$, std=$1.613$) summaries. Similar to study 1, participants with the highest rated familiarity (5 on a 1---5 scale) skipped significantly more sections in the \low{} summaries compared to the \high{} summaries ($d_{unclicked}=0.900$, $p=0.011$, $SMD=0.35$). This estimated difference between skipped sections constitutes close to a full extra section skipped (e.g., skipping all of the summary's limitations). 

\subsubsection{\new{Original article requests}}
\label{sec:resultsTwoOriginalArticle}

\new{Participants on average requested the original article 52.5\% of the time. Generally participants requested the original article from the \low{} summaries more often (mean=55.5\%) than either the \med{} (mean=48.2\%) or \high{} (mean=53.9\%). In contrast to study 1, where low familiarity participants requested the original article more for \high{} summaries, participants in study 2 with the second lowest familiarity requested the original article significantly \textit{more} often in the \low{} summaries compared to the \med{} summaries ($d_{requests}=0.214$, $p=0.036$ $SMD=0.68$). Table \ref{tab:pairwise-contrasts-2} in the appendix lists all pairwise differences.}

\vspace{0.2em}

The results from study 2 corroborate and expand on our findings from study 1. Participants with low familiarity preferred generated low complexity summaries, while high familiarity participants again skipped sections of low complexity summaries more often. \new{One contrasting finding from study 2 was that some participants with low familiarity requested the original article more often for low complexity summaries over more complex summaries.} Given that we observed similar findings from study 1 with expert-written summaries, the results of study 2 suggest that machine-generated summaries are a viable method for efficiently adjusting language to different audiences.

%% file: 05-study3.tex
\section{Study 3 - Machine-Generated Summaries preserving information content}
\label{sec:studyThree}

Summaries from studies 1 and 2 had no restriction on what information needed to be included. This followed past work in plain language summarization, where writers or models select some information to explain, and remove other information (e.g., focusing on a single finding or concept for low complexity text) \citep{guo2022cells, august2020writing, Srikanth2020ElaborativeSC}. However, selectively conveying information comes with the risk of removing information a reader might want \citep{August2022PaperPM}, or giving a reader a false sense of understanding \citep{Scharrer2012TheSO}. Emboldened by newer, stronger models being released (e.g., ChatGPT, or GPT-4), we were curious if generated text could preserve details from high complexity summaries in their low complexity counterparts, potentially mitigating the risk of information loss. This motivates our third research question: 

\begin{quote}
	\textbf{RQ3:} How do participants respond to generated scientific summaries at different complexities if they report similar information?
\end{quote}

\subsection{Method}
\label{sec:studyThreeMethod}
Below we describe our method for generating summaries in study 3. Please refer to \S\ref{sec:methods} for shared methodology of studies 1, 2, and 3.

\subsubsection{Materials}
\label{sec:studyThreeMaterials}

\paragraph{Generating detail-preserving summaries}

In studies 1 and 2, there was no requirement that summaries preserve information (i.e., it was acceptable if a simpler summary removed some information). For study 3, we sought to generate low complexity summaries that preserved information content (i.e., were plainer but included all details). We did this by leveraging stronger models released after study 2 and developing a prompting technique to simplify each sentence separately, prompting the model to elaborate on details rather than remove them. In simplification literature, both removing and elaborating on details are common tasks \citep{laban-etal-2023-swipe, ch2023medeasi}. In the context of study 3, we structured model input and prompts to minimize detail removal and maximize elaboration for all details in the original sentence. We used GPT-3.5 Turbo in May 2023 with temperature set to 1.0 and the rest of the parameters set to default OpenAI API settings.

We generated summaries that did not remove and instead elaborated on details by restricting the model input and changing our prompting technique. Rather than input the entire \high{} summary, as in study 2 (\S \ref{sec:studyTwoMaterials}), we provided GPT-3.5  with a single sentence at a time and instructed it to explain, rather than remove, any information from the original sentence. To avoid having subsequent sentences repeat themselves, the prompt included the history of previous simplified sentences and instructed the model not to explain a concept it had explained above. In addition to the instructions, the prompt included one example of a scientific sentence and its associated simplified version. 

We used two prompts, one for \med{} summaries and one for \low{}. The \med{} prompt instructed the model to rewrite the sentence for someone very familiar with the topic of the sentence, with a target reading level of a college educated adult. For the \low{} summaries the target user was someone who was not at all familiar with the sentence's topic, with a target reading level of 5th grade. 5th grade was chosen based on previous work in generating plain language summaries \citep{August2022PaperPM}, and on our observations that selecting a high school reading level, as we had done for the expert-authored summaries, produced text similar to the \med{} prompt. The full prompts are included in Appendix \ref{app:generatingSummariesStudy3}. Table \ref{tab:exampleSummaries} provides examples of the generated summaries.

\subsubsection{Measuring information content in summaries}
\label{sec:studyThreeInfoEval}

Before collecting participant response to the summaries, we analyzed how information content differed between the summary versions in the three studies. We used four automatic measures and one manual measure of information content based on previous work studying alignment between scientific text and summaries \citep{longeval23, guo2022cells, ernst-etal-2021-summary}: 

\begin{itemize}
    \item[] \textbf{SummaC}: \citet{laban-etal-2022-summac} introduced an natural language inference (NLI) approach to summary consistency. The method uses an NLI model to score each sentence from a source summary with sentences from a target summary on how much the target sentences follow from the source sentence (i.e., is true given the source sentence). We use the \textsc{SummaC-Conv} model using the default settings from the original metric library.\footnote{\url{https://github.com/tingofurro/summac/tree/master}}
    
    \item[]
    
    \item[] \textbf{SuperPAL}: \citet{ernst-etal-2021-summary} introduced a supervised method for scoring alignment between source and target summaries by annotating spans of text representing information units (i.e., a standalone fact). Using these annotated spans, the authors trained a model for the task of identifying information alignment between a source and target summary.  In an evaluation of alignment scores for scientific summaries, SuperPAL was found to be the most effective at identifying aligned claims between the source and target \citep{longeval23}. We use the \textsc{bui-nlp/superpal} model\footnote{\url{https://github.com/martiansideofthemoon/longeval-summarization}} with the default settings.
    
    \item[]
    
    \item[] \textbf{ROUGE-L}~\citep{lin-2004-rouge}: ROUGE is a common score for assessing summary quality by scoring the number of n-gram overlaps between source and target summaries. ROUGE has also been used as a baseline approach to aligning sentences between source and target summaries \citep{guo2022cells}. Following this prior work, we use ROUGE-L, which measures the longest common subsequence of tokens between a source and target sentence. We use the Huggingface \textsc{evaluate} package for calculating ROUGE-L.\footnote{\url{https://github.com/huggingface/evaluate/tree/main}}

    \item[]
        
    \item[] \new{\textbf{BERTScore}~\citep{Zhang2020BERTScore}: BERTScore is a common score for summary evaluation that computes semantic similarity using pre-trained contextual embeddings from the BERT model \citep{Devlin_2019}. We use the Huggingface \textsc{evaluate} package for calculating BERTScore and report the F1 score.\footnote{\url{https://github.com/huggingface/evaluate/tree/main}}}

\end{itemize}

For each measure we take the average maximum alignment score for sentences in the \high{} summaries with sentences from the \med{} and \low{} summaries. If a sentence in the \high{} summary has low alignment scores for all sentences in the \med{} or \low{} summaries, this would suggest that the information is not reported in the summaries.

\begin{table*}[t!]
\centering \small
\begin{tabular}{llrrrrr}
\textbf{Source} & \textbf{Complexity} & \textbf{SummaC} & \textbf{SuperPAL} & \textbf{ROUGE-L } & \textbf{\new{BERTScore}} & \textbf{Info. Units}\\ \hline

\multirow{2}{*}{Expert - Study 1} & \med{} & $0.086_{.188}$  & $0.227_{.012}$ & $0.211_{.115}$ & $0.879_{0.026}$ & $0.557_{.255}$ \\
& \low{} &  $0.098_{.194}$ & $0.225_{.001}$  & $0.197_{.102}$  & $0.878_{0.024}$  &  $0.478_{.241}$  \\ \\

\multirow{2}{*}{Machine - Study 2} & \med{} & $0.782_{.359}$  & $0.673_{.161}$ & $0.730_{.308}$ & $0.957_{0.047}$ & $0.810_{.317}$ \\
& \low{} &  $0.290_{.360}$ & $0.384_{.250}$  & $0.203_{.128}$ & $0.882_{0.027}$ & $0.418_{.264}$ \\ \\

\multirow{2}{*}{Machine - Study 3} & \med{} & $0.839_{.263}$  & $0.722_{.042}$ & $0.439_{.124}$ & $0.926_{0.022}$ & $0.998_{.014}$ \\
& \low{} &  $0.750_{.302}$ & $0.684_{.077}$  & $0.313_{.112}$ & $0.910_{0.022}$ & $0.977_{.062}$ \\ \\

\end{tabular}
\caption{Differences in automated information content measures between summary versions.}
\label{tab:information_results_auto}
\end{table*}

In addition to the automatic measures reported above, we ran a manual evaluation of the information content between each of the summary version. We annotate all information units---defined similar to prior work as proposition-level semantically equivalent statements \citep{ernst-etal-2021-summary}---for the \high{} summaries and count how many of these units appear in the \med{} and \low{} summaries. Annotating information units at this level has been used in prior work for evaluating claims in scientific summaries \citep{longeval23}. In our summaries these units were predominately definitions of terminology, reporting of results, methodological details, and background explanations. Our codes are provided in the supplementary.    

Table \ref{tab:information_results_auto} lists the scores for summaries' information content. Across all measures and versions, the \low{} summaries score lower than the \med{} summaries. The most common information skipped in all the summaries (based on our manual evaluation of information units) was information about the findings from the studies. This aligns with feedback from our writer, who said that in the \low{} summaries they focused on only the most import finding, while in the \med{} summaries they included more details. One reason for the lower scores on most automatic measures for the expert summaries might be due to the writer using fewer overlapping words compared to the models. The same can explain the higher ROUGE-L score for the study 2 \med{} summaries, which used many spans verbatim from the original summaries. In comparison to the summaries from studies 1 and 2, though, the summaries in study 3 have consistently higher scores and differences between the \med{} and \low{} versions are within 1.5 standard deviations.

\subsection{Results}
\label{sec:studyThreeResults}

\subsubsection{Reading experience measures}

\begin{figure*}
     \centering
     \includegraphics[width=\textwidth]{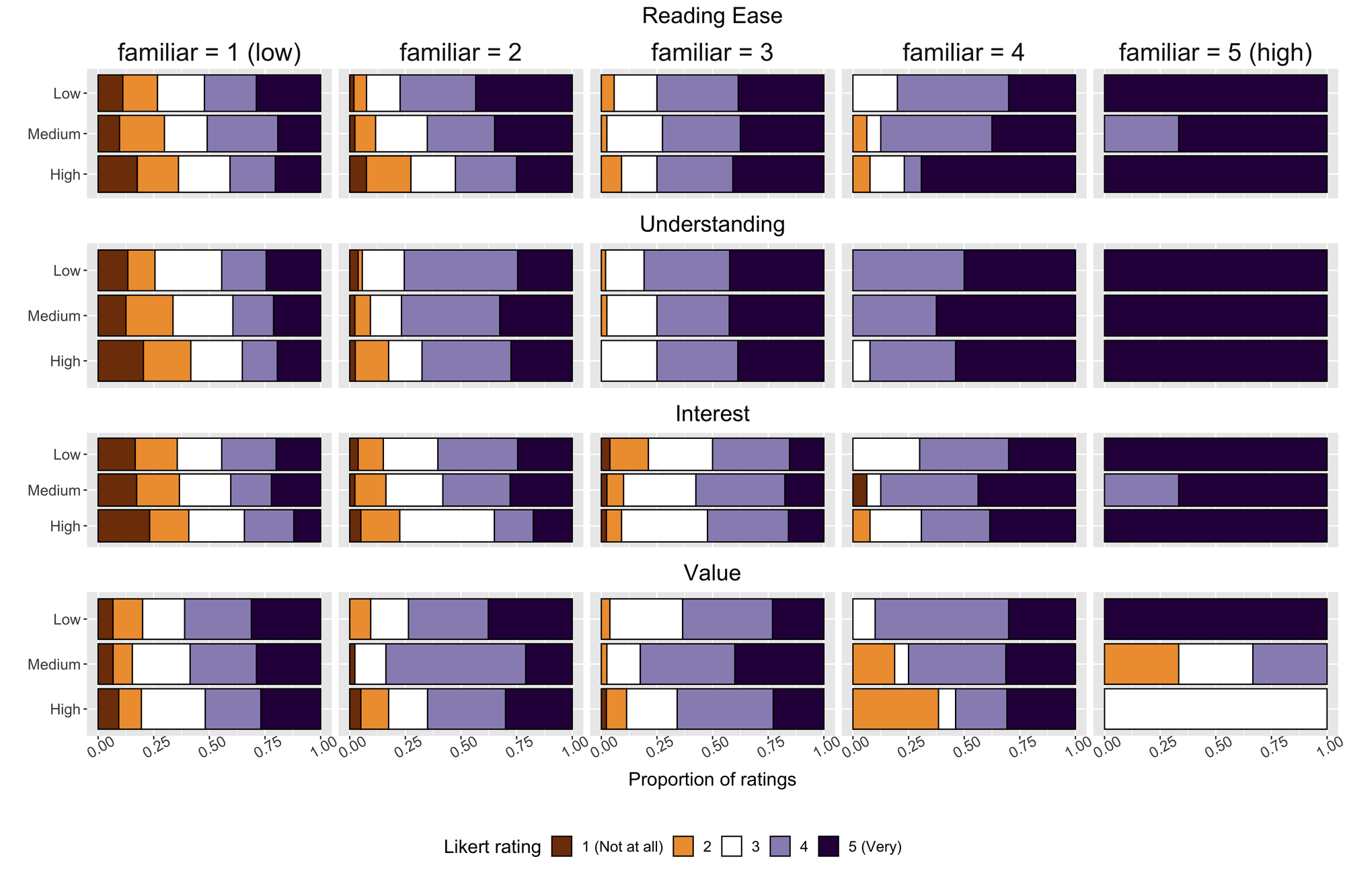}
        \caption{Distribution of ratings  for each reading experience measure across complexity and participant topic familiarity for study 3 (machine generated summaries with information restriction).}
    \label{fig:complexityInteractionsStudy3}
\end{figure*}

Compared to the first two studies, there were smaller differences in reading experience ratings between the three complexity versions. Figure \ref{fig:subjective_ratings_study3} plots the overall ratings. While participants generally rated \low{} summaries as  easier to read ($d_{\textit{ease}}=0.166$, $p=1.0$, $SMD=0.29$) and understand ($d_{\textit{understand}}=0.734$, $p=0.051$, $SMD=0.24$) compared to \high{} summaries, these differences were smaller and not significant.

Participants who had the lowest familiarity of the summary's topic again rated the \low{} summaries as significantly easier to read and understand than the \high{} summaries ($d_{ease}=0.362$, $p=0.019$, $SMD=0.27$; $d_{understand}=0.420$, $p=0.003$, $SMD=0.28$). Similar to studies 1 and 2, participants with more familiarity rated the three summary versions similarly, with no significant differences between them. Figure~\ref{fig:complexityInteractionsStudy3} plots ratings broken down by familiarity.

\subsubsection{Skipped sections}

In Study 3, participants on average skipped $0.554$ (std=$1.220$) sections. Similar to studies 1 and 2, skipped sections were lowest for the \high{} summaries (mean=$0.490$, std=$1.134$) compared to the \low{} (mean=$0.529$, std=$1.209$) and \med{} (mean = $0.642$, std=$1.310$) summaries. Participants who rated their topic familiarity as a 3 out of 5, indicating moderate familiarity, skipped significantly more sections in the \med{} summaries compared to the \high{} summaries ($d_{unclicked}=0.472$, $p=0.026$, $SMD=0.57$) and \low{} summaries ($d_{unclicked}=0.583$, $p=0.004$, $SMD=0.51$).

\subsubsection{\new{Original article requests}}
\label{sec:resultsThreeOriginalArticle}

\new{Similar to study 2, participants requested the original article from the \low{} summaries more often (mean=18.7\%) than either the \med{} (mean=12.8\%) or \high{} summaries (mean=12.8\%). Also supporting our results from study 2, participants in study 3 with the lowest familiarity requested articles significantly more often after reading the \low{} summaries compared to the \med{} summaries ($d_{requests}=0.108$, $p=0.023$, $SMD=0.39$) and \high{} summaries ($d_{requests}=0.110$, $p=0.023$, $SMD=0.34$). Table \ref{tab:pairwise-contrasts-3} in the appendix lists all pairwise differences.}

%% file: 08-discussion.tex
\section{Discussion}
\label{sec:discussion}

In this paper we set out to understand how general audience readers with different background knowledge respond to alternative versions of scientific language. We conducted three studies, using both human-written and machine-generated text, investigating the effect of language complexity and topic familiarity on reading experience and behavior. We found that the lowest complexity summaries, both human-written and machine-generated, provided the most benefit to readers with little familiarity of a scientific topic (e.g., those who had never heard of the summary's topic before). \new{Not only did low complexity summaries make it easier for low familiarity participants to read and understand the summaries, but in the case of machine-generated summaries, the low complexity summaries also encouraged them to request the original scientific article more, engaging with the science beyond what was required for the study.} 

In most cases, though, the benefits of low complexity came at the cost of reduced information content. In our first two studies, low complexity summaries provided less information overall than high complexity summaries, especially in reporting multiple findings (\S\ref{sec:studyThreeInfoEval}). In our third study, when we encouraged models to generate plain language that preserved details, we found that only readers with the lowest topic familiarity rated the longer plain summaries as easier to read and understand (\S\ref{sec:studyThreeResults}). Most science communication text focuses on the most important findings and theories to convey by default \citep{august2020writing, CochranePLS}. This is because reporting all scientific findings in plain language requires explaining any concepts an audience might not know \citep{Wu2023ElaborativeSA}, leading to long explanations that risk reader fatigue and loss of interest. \new{Our findings from study 3 align with this work by showing that conveying complete information in plain language leads to longer summaries that were only easier to read for those who had no background in the summary's topic.}

While lower complexity summaries might be ideal for low familiarity readers, they may invite high familiarity readers to ignore information. Across the three studies, participants with higher familiarity skipped sections of low complexity summaries significantly more than high complexity summaries. This could potentially be due to lack of interest, or feeling like the summary was talking down to them \citep{MartnezSilvagnoli2022OptimizingRA}. In some cases, the difference in number of skipped sections was close to one section out of five. While not all information may be necessary to convey, the skipped information was often the most risky to skip: the study’s limitations.

\new{Our findings are the first to illustrate the benefits and drawbacks of simplification for general audience readers with varying background knowledge. Prior work developing systems to support science communication has predominantly focused on providing a single version of simplified language and treated general audience readers as a single, monolithic group \citep{Guo2023APPLSAM, Devaraj2021ParagraphlevelSO}. While science communicators have a strong intuition that adapting language to different audiences is important \citep{august2020writing}, no work has taken the step of showing that such adaptation can provide measurable benefits. In our three studies, we show that the simplest summaries benefit readers with the least knowledge of a topic the most, and that more complex summaries are best for those with greater background knowledge.}

\subsection{\new{Guidance on adaptive plain language}}
\label{sec:guidanceDiscussion}
This paper provides guidance on designing generated language for both science communicators and interface designers. Based on our findings we make the following suggestions:

\begin{itemize}
    \item \textbf{Low complexity for low familiarity/information} The least complex plain language summaries are better when one or both of the following is true: there is no requirement to convey complete information (\S\ref{sec:studyOne} \& \ref{sec:studyTwo}), or the reader has little to no familiarity in the topic (in this case longer, plain summaries can be used, \S\ref{sec:studyThree}). 
    \item \textbf{High complexity for high familiarity} More complex summaries---even with text drawn from the research paper---are better when audiences have more background knowledge (even if they are not experts in a field), in order to convey more information and keep readers engaged (\S\ref{sec:studyOne} \& \ref{sec:studyTwo}). 
    \item \textbf{Plain language for high information, when necessary} LMs can be used to generate plain language summaries that preserve details (\S\ref{sec:studyThreeMaterials}); however, these summaries only benefit those with little knowledge of a scientific topic (\S\ref{sec:studyThree}) and should be used only when necessary because it leads to much longer text that risks losing readers that have even moderate topic familiarity. 
\end{itemize}

\new{Science communicators can use our findings to guide their efforts when  reaching different audiences. If an article is intended for readers with no familiarity in a topic, a science writer could meet these needs by generating and editing a very plain summary or by assessing their own writing with automatic complexity measures (\S\ref{sec:studyOneSimpleLangEval}). In contrast, if a science communicator is worried about losing the engagement of readers with more topic familiarity, they could focus on a more complex summary, either generated or written. Further, a writer could create multiple alternative versions of a summary suited for different audiences quickly using our generation techniques (\S\ref{sec:studyTwoMaterials} \& \ref{sec:studyThreeMaterials}).}

\new{Interface designers can also leverage the techniques we illustrate in our studies to create interactive and adaptive reading interfaces. For example, a reading interface could generate a new summary on-the-fly based on the reader, or allow readers to interactively select different versions as they read. Short user surveys could be used to determine the ideal adaptation \citep{Wallace2022TowardsIR}, similar to the method employed in this paper (\S\ref{sec:readingMeasures}). A complementary method would be to model users through behavioral signals, a common approach in the education literature \citep{Desmarais2012ARO, Kotseruba201840YO}. We observed that participants with higher familiarity were more likely to skip sections when the complexity was too low. A system that adapts scientific complexity could monitor how much skipping a reader engages in, increasing complexity with increased skipping. Another approach to modeling a user in this context is to analyze past reading or writing behavior \citep{Amith2020MiningHV}. A system could predict an ideal complexity based on the terminology and concepts contained in documents a user already knows. We recommend some level of user control for adaptive language. While users might not always know the ideal level of complexity for themselves, an adaptive language system could also include a knob or dial that allows a reader to scan through possible versions if the current adaption is not ideal. }

One major hurdle in deploying systems using language models is the risk of hallucinations. We argue that such hallucinations necessitate human expert involvement. Rather than expert involvement being a limitation, though, we envision it improving human-human communication across the barriers that scientific language can impose. Science communication is ideally a conversation, not only a transmission of information~\citep{nisbet2009s}. Our hope is that requiring expert oversight will help science communicators quickly create summaries that serve diverse audiences while also encouraging communicators to think deeply about the audiences they are reaching with their work.

\section{Conclusion}

In this paper, we investigate how general audience readers respond to scientific summaries written or generated at different levels of complexity. Across our three studies, using expert-written and machine-generated summaries, we show that the ideal text is based on a participant's familiarity of a topic. Low familiarity participants rated the low complexity summaries as easiest to engage with. High familiarity participants rated the summaries equally regardless of complexity, while skipping more sections of low complexity summaries. We also find that using traditional generation or science communication techniques often leads to loss in information as language becomes less complex, but that new generative models are capable of generating plain text while explaining complex topics, retaining much of the information of higher complexity summaries. Our findings highlight the tradeoffs in adapting language complexity for different audiences and provide a path forward for communicating scientific information to a wider range of people.

%% file: 09-appendix.tex
\section{\new{Automated Complexity Measures}}
\label{app:complexityMeasures}
\new{Below we describe in more details the automated complexity measures used in \S\ref{sec:materials}}.
\begin{itemize}
    \item[] \textbf{Thing Explainer out-of-vocabulary (TE)}: We count the ratio of words outside the top 1,000 most common words in English. The words are based on Wiktionary's contemporary fiction frequency list.\footnote{\url{https://en.wiktionary.org/wiki/Wiktionary:Frequency_lists/Contemporary_fiction}} This method was popularized by the popular book \emph{Thing Explainer}, which explains scientific concepts using only the 1,000 most frequent words in English \citep{munroe_2017}.
    
    \item[]
    
    \item[] \textbf{Function words} In medical communication, the proportion of function words (e.g., prepositions, auxiliary or verbs) was found to be positively correlated with perceived and actual readability \cite{leroy2008evaluating, leroy2010influence}. We measure the proportion of function words in a sentence using \texttt{scispacy} \citep{neumann-etal-2019-scispacy}.
    
    \item[]
    
    \item[] \textbf{Language model perplexity (GPT ppl.)} Language models are systems for predicting words in a sequence. The perplexity of the model is a measure of how different a sequence of text is from the language the model was trained on. Perplexity has been found to correlate with perceived and actual reading difficulty \cite{pitler-nenkova-2008-revisiting, collins2014computational}. We use the GPT model \cite{Radford2018ImprovingLU} to measure language model perplexity, as it was trained on common English (as opposed to scientific text).

\end{itemize}

\section{\new{Ordinal Regression for Likert-Scale Variables}}
\label{app:ordinalRegression}
\begin{table*}[h!]
\centering
\begin{tabular}{rrcccccc}
\toprule 
Measure & Model &
 Study 1 & $p$ &%
 Study 2 & $p$ &%
 Study 3 & $p$ 
 
\\

\midrule              

\multirow{2}{*}{Reading Ease}   %
 & CLMM
 & 168.28 & \textbf{<0.001} &%
  94.39 & \textbf{<0.001} & %
 14.04 & \textbf{<0.01}
\\
& LMM
&  181.04 & \textbf{<0.001} & %
  108.33 & \textbf{<0.001} & %
 15.92 & \textbf{<0.05}
\\ \\

\multirow{2}{*}{Understanding} %
& CLMM 
& 118.23 & \textbf{<0.001}&%
 51.22 & \textbf{<0.001} & %
 9.41 & 0.116
\\
& LMM &
 134.63 & \textbf{<0.001}&%
 55.08 & \textbf{<0.001} & %
 12.39 &\textbf{<0.05}
\\ \\

\multirow{2}{*}{Interest}%
& CLMM &
 57.38 & \textbf{<0.001} &%
 18.22 & \textbf{<0.01} & %
 10.11 & 0.116
\\
& LMM &
 61.64 & \textbf{<0.001}&%
 26.88 & \textbf{<0.001 }& %
 9.32 & 0.107
\\ \\

\multirow{2}{*}{Value}
& CLMM &
 18.66 & \textbf{<0.001} &%
 11.09 & \textbf{<0.05} & %
 7.08 & 0.132
\\
& LMM &
 19.64 & \textbf{<0.001}&%
 10.65 & \textbf{<0.01 }& %
 8.82  & 0.107
\\

\bottomrule

\vspace{0.2ex}

\end{tabular}

\caption{
Likelihood ratio test statistics and $p$-values for likelihood ratio test of cumulative link (CLMM) and linear (LMM) mixed-effects models. \textbf{Significant} values are bolded. $p$-values are adjusted using Holm-correction.   
}
\Description{%
}
\label{tab:clmm}
\end{table*}

\new{As our reading experience measures were measured on a Likert-style scale, the linear mixed effects model (LMM) estimates could be ill-suited for analysis, especially if these measures were not sufficiently normally distributed. As an alternative, we additionally fit analogous cumulative link mixed-effects models (CLMM) from the \textsc{ordinal} R package \cite{Christensen2018CumulativeLM} and conducted likelihood ratio tests, which are similar to 
\emph{F}-tests but more conservative, on the interaction term of complexity level of article familiarity.}

\new{To accurately identify the effect complexity has on our measures and its interaction with topic familiarity, we define two models for each measure. Each model includes the same random effects of paper ID and participant ID to control for variation among papers and participants. }

\begin{enumerate}
    \item $\mathit{LMM}_{\mathit{full}}$: Containing fixed effects for the complexity version, topic familiarity, an interaction term for familiarity and complexity, and random effects for paper and participant IDs.
    \item $\mathit{LMM}_{\mathit{none}}$: Containing a fixed effect for topic familiarity and random effects for paper and participant IDs.
\end{enumerate}

\new{With these models we evaluate how complexity affects reading measures (e.g., reading ease) by comparing the model goodness-of-fit between $\mathit{LMM}_{\mathit{full}}$ and $LMM_{none}$ using the $\chi^2$ likelihood-ratio test. If $\mathit{LMM}_{\mathit{full}}$ has a significantly stronger fit, this suggests that complexity has a significant effect on that reading measure. }

\new{Table \ref{tab:clmm} lists the $p$-values for the likelihood ratio tests on the CLMM and LMM models. The $p$-values are similar across the two methods, with the one exception being a significant difference in understanding for Study 3. To confirm our findings of differences across complexity measures, we additionally ran Mann–Whitney $U$-tests on the reading experience ratings. While the studies were within-subjects, we treated the data as unpaired because familiarity ratings differed across the same participant, and therefore were not grouped together. While these tests did not control for participant or paper random effects (as the post-hoc $t$-tests we report in the results do), the findings remained similar to those reported in Tables \ref{tab:pairwise-contrasts-1}, \ref{tab:pairwise-contrasts-2} and \ref{tab:pairwise-contrasts-3}. Following prior work \citep{August2022PaperPM, Head2021AugmentingSP, Norman2010LikertSL}, we report results from the parametric tests (i.e., LMMs and pairwise difference $t$-tests) in the paper.}

\begin{table*}[]
\centering 
\begin{tabular}{p{20mm}p{120mm}}
\toprule
 \textbf{Prompt Level} & \textbf{Summary}   \\ \midrule \midrule 
    Second-grade student & In this study, people who could do more than 40 pushups were associated with a reduced risk of having a heart problem in the next 10 years, compared to those who could do fewer than 10. The study also found that people who could do more than 10 pushups had a reduced risk of having a heart problem, and that the risk decreased as the number of pushups increased.  \\ \\
    
    Seventh-grade student & This study found that people who could do more than 40 pushups were less likely to have a heart problem in the next 10 years. The study also found that the more pushups someone could do, the less likely they were to have a heart problem.  \\ \\

    College student & This study found that people who could do more than 40 pushups had a significantly lower risk of having a cardiovascular disease event, compared to those who could do 10 or fewer. This may be because those who could do more pushups also had better baseline cardiovascular disease risk factors. They also found that people who could do 11 or more pushups at the start of the study had a lower risk of having a cardiovascular disease event during the study. \\ \\

\bottomrule

\end{tabular}
\caption{Examples of the generated summaries with different prompts for study 2 using GPT-3. Note that the prompts were not used to select complexity levels. This part of the summaries was under the heading ``What did the paper find?'' Notice how the second grade prompt is slightly longer and uses larger words (e.g., ``associated with reduced risk'' compared to ``less likely to'') than the seventh grade prompt. At the same time, the college student prompt uses more complex language (e.g., ``cardiovascular disease event'') compared to both other generations.}
\label{tab:examplePromptedSummaries}
\end{table*}

\section{Generating summaries - Study 2}
\label{app:generatingSummaries}

GPT-3 was not designed to explicitly vary text complexity, so while generations might vary naturally in complexity due to the changes in prompt, there is no guarantee that prompts will align with complexity (i.e.,  prompting GPT-3 with ``Summarize for a first grade student'' will not necessarily lead to lower complexity than prompting with ``tenth grade student''). In a preliminary analysis of the summaries, we found that the summaries, while tending toward simpler with lower grades, could still be quite complex in the first grade prompted version and much simpler at higher grade levels. Table~\ref{tab:examplePromptedSummaries} provides examples of generations and associated prompts.

There are automatic methods for scientific information extraction  \citep{Cohan2019PretrainedLM} and PDF parsing \citep{Lopez2009GROBIDCA, Shen2021IncorporatingVL} that could in the future be used extract information directly from a research paper PDF. We leave such extensions to future work, as our goal was to explore the feasibility of automatically adjusting language complexity. Any errors introduced by other automated methods (e.g., incorrect text from PDF parsing) could muddy our ability to identify how alternate complexity levels perform in our envisioned context.

\section{Generating summaries - Study 3}
\label{app:generatingSummariesStudy3}

The full prompts were: 

\begin{itemize}
    \item \textbf{\low{}:} You are a helpful assistant who will rewrite 5-10 scientific sentences for a reader who is not at all familiar with the sentence's topic. You will be given one sentence after another. For each sentence, define any necessary terms and provide any background knowledge that a reader who is not at all familiar with this topic might need. Your target grade level is 5th grade. You do not need to explain a term or concept you have already explained. Your goal is to help a lay reader understand all the information contained in the original sentence. Below is an example. 
    
    Input sentence: The accretion of terrestrial planets from the solar nebula has been a subject of extensive theoretical modeling and simulation. 
    
    Simplified output: The growth of rocky planets like the Earth by adding particles, called accretion, from the cloud of dust and gas surrounding the early Sun, called the solar nebula, has been studied a lot using theoretical calculations and computer simulations. Theoretical calculations refer to math that is used to make predictions about how things happen based on a particular theory. Computer simulations are programs that model how a system works.

    \item \textbf{\med{}:} You are a helpful assistant who will rewrite 5-10 scientific sentences for a reader who is very familiar with the sentence's topic. You will be given one sentence after another. For each sentence, define any necessary terms and provide any background knowledge that a reader who is very familiar with this topic might need. Your target grade level is a college-educated adult. You do not need to explain a term or concept you have already explained or that the reader is likely to know. Your goal is to help the reader understand all the information contained in the original sentence. Below is an example. 
    
    Input sentence: The accretion of terrestrial planets from the solar nebula has been a subject of extensive theoretical modeling and simulation. 
    
    Simplified output: The formation of terrestrial planets through accumulating dust, gas, and debris, called accretion, from the solar nebula, has been studied extensively using theoretical calculations and computer simulations.
\end{itemize}

\section{Factuality in Generated Summaries}
\label{app:factuality}

Out of 120 generated summaries in study 2 (6 sections $\times$ 10 papers $\times$ 2 complexities), 22 were labelled as containing any hallucinated content. The labels were mutually exclusive. There were three types of hallucinations we identified: correct information not from the original text, incorrect information not from the original text and reversing the direction of findings. Table ~\ref{tab:hallucinations} includes examples of these three hallucinations. 

The extent and kind of hallucinations in our summaries can tell us what risk such hallucinations pose and how much effort an expert must invest to make the summaries publishable. For example, if the majority of hallucinations are new but correct information (a common type of hallucination \citep{Cao2022HallucinatedBF}), then they pose less of a risk and require less expert knowledge to fix than if the hallucinations instead reverse the direction of a found effect (another type of hallucination ~\citep{Devaraj2022EvaluatingFI}). We generated summaries with no restriction on hallucinated content. After generation, one author labelled all generations for hallucinated content.

\begin{table*}[]
\centering 
\begin{tabular}
{p{35mm}p{55mm}p{30mm}p{20mm}}
\toprule
 \textbf{Hallucination type} & \textbf{Example}  &  \textbf{Reason} & \textbf{\% Generations}   \\ \midrule  
 \\
   Incorrect additional information & The study found that the babies of women who ate nuts during pregnancy were less likely to have certain health problems. & Nothing in study about health problems & 7.5\%
   \\ \\
  Correct additional information & These cells work together to make sure that we feel pain when we are hurt. This is important because it helps us to avoid getting hurt again. & Nothing in original article about the importance of pain sensation & 2.5\%
   \\ \\
  Reverse direction of findings & This study found that spending more time playing video games can lead to more aggressive behavior. & Finding was that time spent playing video games did not lead to more aggressive behavior & 4.2\%
   \\
  \\[-5pt]
\bottomrule

\end{tabular}
\caption{Three types of hallucinations encountered in our generated summaries in study 2 (with GPT-3).
}\label{tab:hallucinations}
\end{table*}

Including correct information not from the original text occurred in 3 hallucinations. Usually these hallucinations included text about the study findings with no associated text from the original source text, or else hallucinated the existence of graphs from additional studies (e.g., ``This chart shows the probation rates of the US population ...''). These hallucinations reported correct information, even though the information was not reported in the source text.

9 hallucinations included incorrect information not from the original text. These hallucinations added unrelated findings to the summary that were not reported in the study. Examples include hallucinating an association between asthma and nut intake, while the original article reported on nut intake and neuropsychological development. 

Including correct and incorrect information not from the original text are similar to \textit{extrinsic} hallucinations in the summarization literature~\citep{Goyal2021AnnotatingAM}, or \textit{information insertion} in the simplification literature~\citep{Devaraj2022EvaluatingFI}. Both refer to hallucinations adding information not found in the original source.

Reversing the direction of findings occurred in 5 hallucinations. These hallucinations reported the exact opposite result than was reported in the original study. These hallucinations are considered \textit{intrinsic} hallucinations, or \textit{information substitution} which are hallucinations that include information in direct contrast to the original source~\citep{Maynez2020OnFA, Devaraj2022EvaluatingFI}. 

These three types of hallucinations are well-documented in literature studying generative model hallucinations \citep{Maynez2020OnFA, Goyal2021AnnotatingAM, Cao2022HallucinatedBF, Devaraj2022EvaluatingFI}. We add to this previous literature by showing how such hallucinations occur in this reading context.

We also explored using automated methods to identify hallucinations. We tried two commonly used automated measures for hallucinations, SummaQA \citep{Scialom2019AnswersUU} and entity-level F1~\citep{Nan2021EntitylevelFC}. SummaQA uses a BERT-based question answering model to answer questions extracted from the source text with the summary text. We use the original extracted sentences as the source text. Entity-level F1 measures the number of entities that occur in a generated summary compared to the ground truth summary. We use \texttt{scispacy} \citep{neumann-etal-2019-scispacy} to extract entities. We observed no significant differences in either score between generated summaries with or without hallucinations (two-sided $t$-test $t_{118}=0.04$, $p=0.972$ for SummaQA F-score, $t_{118}=1.90$, $p=0.119$ for entity-level F1 after Holm correction). When inspecting the scores of generations, we also observed that both scores skewed positively (i.e., measured less hallucinated content) towards summaries that had language more similar to the original. This led to the scores negatively impacting the lower complexity summaries since they used language more distinct from the original researcher version. Based on these results, we did not use any automated factuality scores to curate the summaries.

\section{\new{Pairwise test statistics}}
\label{app:testStats}

\new{Below we report all test statistics for pairwise comparisons in the three studies. }

\begin{table*}[h]

\centering
    \begin{tabular}{rrcccccc}
    \toprule & Familiarity & 
    $d^{Lo-Me}$ &%
    $p$ & %
    $d^{Lo-Hi}$ &%
    $p$ & %
    $d^{Me - Hi}$ &%
    $p$  \\
    
    \midrule              
        
    \multirow{5}{*}{Reading Ease} & 1 & 0.554 & \textbf{<0.0001} & 1.490 & \textbf{<0.0001}  & 0.936 & \textbf{<0.0001}  \\
    & 2 & 0.103 & 0.621  & 0.782 & \textbf{0.001}  & 0.679 & \textbf{0.003}  \\
    & 3 & 0.197 & 0.391  & 0.695 & \textbf{0.013}  & 0.498 & 0.059  \\
    & 4 & 0.101 & 0.817 & 0.609 & 0.544 & 0.508 & 0.588  \\
    & All & 0.238 & 0.069  & 0.894 & \textbf{<0.0001} & 0.655 &  \textbf{<0.0001}  \\
    
    \midrule
    
    \multirow{5}{*}{Understanding} & 1 & 0.458 & \textbf{<0.0001} & 1.160 & \textbf{<0.0001} & 0.701 & \textbf{<0.0001} \\
    & 2 & 0.022 & 0.910 & 0.693 & \textbf{0.002} & 0.671 & \textbf{0.002} \\
    & 3 & 0.172 & 0.597 & 0.391 & 0.240 & 0.219 & 0.597 \\
    & 4 & 0.160 & 1.0 & 0.127 & 1.0 & -0.033 & 1.0 \\
    & All & 0.203 & 0.094 & 0.593 & \textbf{<0.0001} & 0.390 & \textbf{0.006} \\
    
    \midrule
    
    \multirow{5}{*}{Interest} & 1 & 0.296 & \textbf{0.021} & 0.943 & \textbf{<0.0001} & 0.647 & \textbf{<0.0001} \\
    & 2 & -0.007 & 0.975 & 0.298 & 0.593 & 0.305 & 0.593 \\
    & 3 & 0.024 & 1.0 & -0.009 & 1.0 & -0.033 & 1.0 \\
    & 4 & 0.864 & 0.220 & 0.261 & 0.603 & -0.603 & 0.520 \\
    & All & 0.294 & 0.085 & 0.373 & \textbf{0.042} & 0.079 & 0.613 \\
    
    \midrule
    
    \multirow{5}{*}{Value} & 1 & 0.314 & \textbf{0.020} & 0.509 & \textbf{<0.0001} & 0.195 & 0.104 \\
    & 2 & -0.012 & 1.0 & 0.009 & 1.0 & 0.021 & 1.0 \\
    & 3 & -0.087 & 1.0 & -0.099 & 1.0 & -0.012 & 1.0 \\
    & 4 & 0.329 & 1.0 & -0.123 & 1.0 & -0.451 & 1.0 \\
    & All & 0.136 & 0.996 & 0.074 & 1.00 & -0.062 & 1.00 \\ 

    \midrule
    
    \multirow{5}{*}{Skipped Sections} & 1 & 0.041 & 0.994 & 0.051 & 0.994 & 0.009 & 0.994 \\
    & 2 & -0.107 & 0.813 & -0.007 & 0.941 & 0.099 & 0.813 \\
    & 3 & -0.252 & 0.056 & -0.008 & 0.943 & 0.244 & 0.056 \\
    & 4 & 0.285 & 0.202 & 0.682 & \textbf{0.008} & 0.398 & 0.202 \\
    & All & -0.008 & 0.892 & 0.179 & \textbf{0.020} & 0.188 & \textbf{0.020}\\

    \midrule
    
    \multirow{5}{*}{Article Requests} & 1 & 0.009 & 0.768 & 0.056 & 0.206 & 0.047 & 0.270 \\
    & 2 & -0.026 & 0.659 & -0.184 & \textbf{0.007 }& -0.159 & \textbf{0.017} \\
    & 3 & -0.078 & 0.439 & 0.027 & 0.685 & 0.105 & 0.287 \\
    & 4 & 0.063 & 1.0 & 0.018 & 1.0 & -0.045 & 1.0 \\
    & All & -0.008 & 1.0 & -0.021 & 1.0 & -0.013 & 1.0\\
            
    \bottomrule
    \vspace{0.2ex}
    
    \end{tabular}
    
    \caption{Post-hoc (two-sided) tests for pairwise differences in fixed-effects estimates between complexity versions and across all participant topic familiarities for study 1 with expert-written summaries. `All' topic familiarity refers to pairwise differences across complexity levels without a topic familiarity subgroup (e.g., average difference across complexity levels.) This table reports the difference in fixed-effects estimates $i - j$ and Holm-Bonferroni-corrected $p$-values~\cite{Holm1979ASS} under our mixed-effects model, where $i$ and $j$ correspond to complexity options. --- $Lo=$ \low{}, $Me =$ \med{}, and $Hi =$ \high{}. Statistically significant $p$-values are bold. For example, in Table~\ref{tab:pairwise-contrasts-1} in the column for $d^{Lo - Hi}$ and row for ``Reading Ease,'' and ``1'' in topic familiarity we can interpret the result as participants with a 1 topic familiarity rated the \low{} complexity, on average, 1.490 points higher for reading ease (out of 5) compared to the \high{} complexity when controlling for participant and paper.}
    \label{tab:pairwise-contrasts-1}
\end{table*}

\begin{table*}[h]
\centering
    \begin{tabular}{rrcccccc}
    \toprule & Familiarity & 
    $d^{Lo-Me}$ &%
    $p$ &%
    $d^{Lo-Hi}$ &%
    $p$ &%
    $d^{Me - Hi}$ &%
    $p$ \\
    
    \midrule              
        
    \multirow{6}{*}{Reading Ease} & 1 & 1.385 & \textbf{<0.0001} & 1.645 & \textbf{<0.0001} & 0.260 & 0.120 \\
    & 2 & 0.310 & 0.274 & 0.660 & \textbf{0.024} & 0.350 & 0.274 \\
    & 3 & 0.392 & 0.101 & 0.321 & 0.161 & -0.071 & 0.683 \\
    & 4 & 0.057 & 1.0 & -0.045 & 1.0 & -0.102 & 1.0 \\
    & 5 & 0.216 & 1.0 & 0.093 & 1.0 & -0.122 & 1.0 \\
    & All & 0.472 & \textbf{<0.0001} & 0.535 & \textbf{<0.0001} & 0.063 & 0.455 \\
    
    \midrule
    
    \multirow{6}{*}{Understanding} & 1 & 0.836 & \textbf{<0.0001} & 1.103 & \textbf{<0.0001} & 0.267 & 0.110 \\
    & 2 & 0.369 & 0.267 & 0.630 & \textbf{0.035} & 0.262 & 0.269 \\
    & 3 & 0.035 & 0.850 & 0.223 & 0.678 & 0.188 & 0.678 \\
    & 4 & 0.030 & 1.0 & -0.077 & 1.0 & -0.107 & 1.0 \\
    & 5 & 0.127 & 0.702 & -0.266 & 0.702 & -0.394 & 0.514 \\
    & All & 0.279 & \textbf{0.004} & 0.323 & \textbf{0.001} & 0.043 & 0.622  \\
    
    \midrule
    
    \multirow{6}{*}{Interest} & 1 & 0.590 & \textbf{0.001} & 0.909 & \textbf{<0.0001} & 0.319 & 0.055 \\
    & 2 & 0.134 & 1.0 & 0.125 & 1.0 & -0.009 & 1.0 \\
    & 3 & -0.047 & 1.0 & -0.064 & 1.0 & -0.018 & 1.0 \\
    & 4 & -0.004 & 1.0 & -0.009 & 1.0 & -0.006 & 1.0 \\
    & 5 & 0.251 & 1.0 & 0.052 & 1.0 & -0.199 & 1.0 \\
    & All & 0.185 & 0.077 & 0.202 & 0.077 & 0.017 &  0.818\\
    
    \midrule
    
    \multirow{6}{*}{Value} & 1 & 0.069 & 0.674 & 0.407 & \textbf{0.031} & 0.339 & 0.079 \\
    & 2 & -0.220 & 0.970 & 0.009 & 0.970 & 0.229 & 0.970 \\
    & 3 & -0.173 & 1.0 & -0.161 & 1.0 & 0.012 & 1.0 \\
    & 4 & 0.151 & 0.665 & -0.085 & 0.665 & -0.236 & 0.427 \\
    & 5 & 0.270 & 0.616 & -0.101 & 0.720 & -0.371 & 0.570 \\
    & All & 0.020 & 1.0 & 0.014 & 1.0 & -0.006 & 1.0 \\

    \midrule
    
    \multirow{6}{*}{Skipped Sections} & 1 & 0.083 & 1.0 & -0.058 & 1.0 & -0.142 & 1.0 \\
    & 2 & -0.333 & 0.645 & -0.143 & 0.924 & 0.190 & 0.924 \\
    & 3 & 0.100 & 1.0 & -0.006 & 1.0 & -0.106 & 1.0 \\
    & 4 & 0.367 & 0.123 & 0.277 & 0.234 & -0.090 & 0.631 \\
    & 5 & 0.235 & 0.424 & 0.900 & \textbf{0.011} & 0.665 & 0.066 \\
    & All & 0.090 & 0.553 & 0.194 & 0.138 & 0.103 & 0.553 \\

    \midrule
    
    \multirow{6}{*}{Article Requests} & 1 & 0.095 & 0.299 & 0.057 & 0.603 & -0.038 & 0.603 \\
    & 2 & 0.214 & \textbf{0.036} & 0.001 & 0.991 & -0.213 & 0.036 \\
    & 3 & 0.001 & 1.0 & -0.046 & 1.0 & -0.047 & 1.0 \\
    & 4 & 0.048 & 1.0 & 0.006 & 1.0 & -0.042 & 1.0 \\
    & 5 & 0.069 & 1.0 & 0.042 & 1.0 & -0.026 & 1.0 \\
    & All &  0.085 & 0.015 & 0.012 & 0.699 & -0.073 & \textbf{0.035} \\

    \bottomrule
    \vspace{0.2ex}

    \end{tabular}
    
    \caption{Study 2 with machine-generated summaries and no restriction on information content. See Table \ref{tab:pairwise-contrasts-1} for examples of pairwise comparison interpretation.}
    \label{tab:pairwise-contrasts-2}
\end{table*}

\begin{table*}[h]
\centering
    \begin{tabular}{rrcccccc}
    \toprule & Familiarity & 
    $d^{Lo-Me}$ &%
    $p$ &%
    $d^{Lo-Hi}$ &%
    $p$ &%
    $d^{Me - Hi}$ &%
    $p$ \\
    
    \midrule              
        
    \multirow{4}{*}{Reading Ease} & 1 & 0.149 & 0.260 & 0.362 & \textbf{0.019} & 0.213 & 0.182 \\
    & 2 & 0.340 & 0.165 & 0.669 & \textbf{0.002} & 0.330 & 0.165 \\
    & 3 & -0.134 & 1.0 & -0.075 & 1.0 & 0.059 & 1.0 \\
    & 4 & 0.235 & 1.0 & 0.201 & 1.0 & -0.034 & 1.0 \\
    & 5 & 0.319 & 1.0 & -0.327 & 1.0 & -0.646 & 1.0 \\
    & All & 0.182 & 1.0 & 0.166 & 1.0 & -0.016 & 1.0 \\

    \midrule
    
    \multirow{4}{*}{Understanding} & 1 & 0.186 & 0.147 & 0.420 & \textbf{0.003} & 0.234 & 0.111 \\
    & 2 & 0.033 & 1.0 & 0.174 & 1.0 & 0.141 & 1.0 \\
    & 3 & -0.169 & 1.0 & -0.062 & 1.0 & 0.107 & 1.0 \\
    & 4 & 0.298 & 0.859 & 0.523 & 0.527 & 0.225 & 0.859 \\
    & 5 & -0.228 & 1.0 & -0.456 & 1.0 & -0.228 & 1.0 \\
    & All & 0.024 & 1.0 & 0.120 & 1.0 & 0.096 & 1.0 \\

    \midrule
    
    \multirow{4}{*}{Interest} & 1 & -0.018 & 0.902 & 0.295 & 0.091 & 0.313 & 0.080 \\
    & 2 & 0.141 & 0.777 & 0.342 & 0.373 & 0.201 & 0.777 \\
    & 3 & -0.291 & 0.613 & -0.173 & 0.853 & 0.117 & 0.853 \\
    & 4 & -0.162 & 1.0 & 0.103 & 1.0 & 0.265 & 1.0 \\
    & 5 & 0.834 & 1.0 & 0.383 & 1.0 & -0.450 & 1.0 \\
    & All & 0.101 & 1.0 & 0.190 & 1.0 & 0.089 & 1.0 \\

    \midrule
    
    \multirow{4}{*}{Value} & 1 & -0.014 & 0.922 & 0.213 & 0.269 & 0.226 & 0.269 \\
    & 2 & -0.025 & 1.0 & 0.165 & 1.0 & 0.190 & 1.0 \\
    & 3 & -0.380 & 0.180 & 0.037 & 0.856 & 0.417 & 0.180 \\
    & 4 & 0.494 & 0.454 & 0.870 & 0.116 & 0.376 & 0.454 \\
    & 5 & 2.245 & 0.177 & 2.385 & 0.200 & 0.139 & 0.906 \\
    & All & 0.464 & 0.139 & 0.734 & 0.051 & 0.270 & 0.284 \\

     \midrule
    
    \multirow{6}{*}{Skipped Sections}& 1 & 0.023 & 1.0 & 0.110 & 1.0 & 0.087 & 1.0 \\
    & 2 & -0.062 & 1.0 & -0.015 & 1.0 & 0.047 & 1.0 \\
    & 3 & -0.583 & \textbf{0.004} & -0.110 & 0.529 & 0.472 & \textbf{0.026} \\
    & 4 & 0.167 & 1.0 & 0.074 & 1.0 & -0.093 & 1.0 \\
    & 5 & -0.055 & 1.0 & -0.203 & 1.0 & -0.148 & 1.0 \\
    & All & -0.102 & 1.0 & -0.029 & 1.0 & 0.073 & 1.0 \\

    \midrule
    
     \multirow{6}{*}{Article Requests} & 1 & 0.108 & \textbf{0.023} & 0.110 & \textbf{0.023} & 0.002 & 0.963 \\
    & 2 & -0.015 & 0.805 & -0.079 & 0.618 & -0.064 & 0.652 \\
    & 3 & 0.101 & 0.347 & 0.082 & 0.352 & -0.018 & 0.783 \\
    & 4 & -0.135 & 0.683 & -0.000 & 1.0 & 0.135 & 0.683 \\
    & 5 & -0.100 & 1.0 & 0.039 & 1.0 & 0.138 & 1.0 \\
    & All & -0.008 & 1.0 & 0.030 & 1.0 & 0.039 & 1.0 \\

    \bottomrule
    \vspace{0.2ex}

    \end{tabular}
    
    \caption{Study 3 with machine-generated summaries and restriction on information content. See Table \ref{tab:pairwise-contrasts-1} for examples of pairwise comparison interpretation.}
    \label{tab:pairwise-contrasts-3}
\end{table*}